\documentclass[%
 reprint,
 amsmath,amssymb,
 aps,
]{revtex4-2}

\usepackage{graphicx}% Include figure files
\usepackage{dcolumn}% Align table columns on decimal point
\usepackage{bm}% bold math
\usepackage{lineno}

\usepackage[utf8]{inputenc}
\usepackage[T1]{fontenc}
\usepackage{booktabs, array, mathptmx, float, tabularx, booktabs, lipsum, amsmath,multirow}
\usepackage{siunitx, xcolor}
\usepackage[version=4]{mhchem}
\graphicspath{{figs/}{figsgaoerb/}} % 读取图片从figs/文件夹中读取，而不是当前文件夹
\usepackage[colorlinks,linkcolor=blue,anchorcolor=blue,citecolor=blue]{hyperref}
\usepackage{braket}
\usepackage{footnote}
\usepackage{mathrsfs}
\usepackage{amsmath}
\usepackage{amssymb}
\usepackage{wasysym}
\usepackage{color,soul}
\usepackage{physics}
\usepackage{siunitx}
\usepackage{dsfont}
\usepackage{float}
\usepackage[english]{babel}
\usepackage{blindtext}
\usepackage[english,nomargin,inline,marginclue,draft]{fixme}
\pdfpageheight\paperheight
\pdfpagewidth\paperwidth
        
\fxusetheme{colorsig}
\FXRegisterAuthor{cg}{acg}{CG}  
\FXRegisterAuthor{th}{ath}{\color{blue}TH}  
\FXRegisterAuthor{ib}{aib}{\color{red}IB} 
\FXRegisterAuthor{sh}{ash}{\color{cyan}SH} 
\FXRegisterAuthor{db}{adb}{\color{green}DB} % now I can ..
\FXRegisterAuthor{ps}{aps}{PS}
\makeatletter
\renewcommand*\FXLayoutInline[3]{%
  {\@fxuseface{inline}\ignorespaces{\color{fx#1}[#3: #2]}}}
\makeatother

\long\def\symbolfootnote[#1]#2{\begingroup%
\def\thefootnote{\fnsymbol{footnote}}\footnotetext[#1]{#2}\endgroup}

\def\nobreakbefore{%
  \relax\ifvmode\else
    \ifhmode
      \ifdim\lastskip > 0pt\relax
        \unskip\nobreakspace
      \else % added to put a ~if no space was typed. (Unclear why it sometimes worked before )
        \nobreakspace
      \fi
    \fi
  \fi
}
\let\oldcite\cite
\renewcommand\cite{\nobreakbefore\oldcite}
\begin{document}

\title{Generation of Near-ideal Indistinguishable Two-Photon State by Incoherent Light}% Force line breaks with \\

\author{Yue-Wei Song$^{1,2}$}
\author{Ming-Yuan Gao$^{1,2}$}
\author{Zhi-Cheng Guo$^{1,2,3}$}
\author{Zheng-He Zhou$^{1,2}$}
\author{Yin-Hai Li$^{1,2}$}
\author{Guang-Can Guo$^{1,2,3}$}
\author{Zhi-Yuan Zhou$^{1,2,3}$}
\altaffiliation {Corresponding author: zyzhouphy@ustc.edu.cn}%Lines break automatically or can be forced with \\
\author{Bao-Sen Shi$^{1,2,3}$}%
\email{Corresponding author: drshi@ustc.edu.cn}

\address{{$^1$}CAS Key Laboratory of Quantum Information, University of Science and Technology of China, Hefei, Anhui 230026, China\\
{$^2$}CAS Center for Excellence in Quantum Information and Quantum Physics, University of Science and Technology of China, Hefei 230026, China\\
{$^3$}Hefei National Laboratory, University of Science and Technology of China, Hefei 230088, China}

%\address {$^1$CAS Key Laboratory of Quantum Information, University of Science and Technology of China, Hefei, Anhui 230026, China}

%\address {$^2$CAS Center for Excellence in Quantum Information and Quantum Physics, University of Science and Technology of China, Hefei 230026, China}

%\address {$^3$Hefei National Laboratory, University of Science and Technology of China, Hefei 230088, China}

\date{\today}% It is always \today, today,

\begin{abstract}
High-quality quantum states lie at the heart of advanced quantum information processing. The degree of photon indistinguishability is critical for applications from photonic quantum computation to precision metrology. The two-photon Hong–Ou–Mandel (HOM) interference effect provides a rigorous quantification method, with its visibility serving as the ultimate benchmark for source quality. Generally, the coherent pumping is widely regarded as indispensable for the preparation of quantum sources. As a result, incoherent light sources have seen limited applications in the current quantum technologies. In this work, we generate an indistinguishable two-photon state by incoherent light generated by frequency doubling of Amplified Spontaneous Emission light. The theoretical analysis indicates that phase randomization of the pumping does not affect the coincidence visibility in two-photon intensity interference. Moreover, temporal incoherence further enhances the symmetry of the generated spectrum in second-harmonic generation. In the experiment, the incoherently pumped photon sources exhibit a heralding efficiency of approximately 60\% and a coincidence-to-accidental ratio exceeding 15000. The observed HOM interference fringes show the visibility of 99.1\% without any spectrum filtering, confirming the near-ideal indistinguishability of the photons. Our study reveals the role of temporal coherence in second-order nonlinear interactions, it provide a potential approach to use an easily accessible incoherent light for engineering high-quality quantum sources.

\end{abstract}

\maketitle

%\tableofcontents

\emph{Introduction}--For decades, advances in quantum photonics have significantly driven the development of quantum information science. Quantum states serving as the physical information carriers are used to implement fundamental operations in these systems\cite{1,2,3,4,5,6,7}. As the foundational resource, the quality of quantum states directly determines overall performance of the system, of which photon indistinguishability stands out as a critical metric\cite{8,9}. When two indistinguishable photons arrive simultaneously at a balanced (50:50) beam splitter (BS), they bunch and exit through the same port, which is the effect known as the Hong–Ou–Mandel (HOM) interference\cite{10,11,12,13}. This quantum property renders the HOM interferometer indispensable for quantum information processing: it facilitates the implementation of logical gates in linear-optical quantum computing\cite{14,15,16}, reduces side-channel information leakage from the source in measurement-device-independent quantum key distribution\cite{17,18,19}, and provides attosecond-resolution essential for high-precision optical metrology\cite{20,21}.

Generally, indistinguishable photons are generated via the nonlinear optical response of media, for example through Spontaneously Parametric Down-Conversion (SPDC) in a nonlinear crystal, in which a single pumping photon is annihilated and a signal–idler photon pair is created simultaneously\cite{22,23,24}. Usually, the coherence of the pumping beam is universally acknowledged as a principal factor governing the quality of the generated photon sources. However, coherent-light pumping imposes stringent requirements on phase stability and timing synchronization, thereby increasing system complexity and constraining scalability in practical implementations\cite{56}. Intrinsic decoherence from laser linewidth broadening fundamentally limits coherent light schemes in long-distance quantum communication and sensing, as quantum noise accumulates once propagation times exceed the laser’s coherence time\cite{29,30,31}. 

Accordingly, there is growing interest in exploiting naturally incoherent or partially coherent light sources for quantum resource generation\cite{32,33,34,35}, such sources are already well established in imaging, optical computing, and remote sensing\cite{36,37,38,39,40,41}. Recently, considerable progress has elucidated the role of coherence in both classical and quantum nonlinear processes. With significant achievements in theoretical formulations and experimental demonstrations, the incoherent light pumped quantum sources have shown great potential\cite{42,43,44}.  Although considerable progress has been made in quantum states generation, previous studies have focused primarily on the role of spatial incoherence. However, spatial incoherence introduces unavoidable losses during states preparation and detection, severely restricting practical applicability\cite{57}. Moreover, the spectral indistinguishability of incoherently pumped states also remains to be demonstrated, the shortcomings hinder the broader deployment of the sources.

To address the challenges, we employ Amplified Spontaneous Emission (ASE) light to generate quantum states\cite{45,46,47}. The inherently uncorrelated spectral components make it an ideal source for investigating temporal incoherence, while its high spatial coherence ensures efficient system performance. Here, both classical and quantum nonlinear processes under incoherent pumping are investigated. We examine the effect of the temporal incoherence of the pumping on the indistinguishability of the generated photon pairs through a HOM interferometer. Theoretical results show that HOM interference relies on the exchange symmetry of joint spectral amplitudes, random distribution of pumping phase cancels out in second-order intensity correlations, so only indistinguishability matters. Simultaneously, the phase-uncorrelation leads to the intensity convolution of incoherent pumping in second harmonic generation (SHG) process. This effect inherently bolstering resistance to spectral perturbations. Compared to coherent pumping, it produces a markedly more symmetric output spectrum.

Experimentally, ASE is used to probe the behavior of incoherent light in ${\chi ^{(2)}}$ processes. We demonstrate the generation of highly symmetrical spectra through SHG. Besides, we produce the quantum photon source via SPDC and characterize the quantum correlation properties. Through HOM interferometer, we observe high-visibility interference fringes, confirming the spectral indistinguishability of the incoherently pumped photons. Finally, the relationship between the purity and Fisher Information are discussed based on the bandwidth of HOM fringe.

\emph{Theoretical Analysis}--The origin of the HOM effect arises from the quantum interference of two indistinguishable single photons impinging simultaneously on a 50:50 BS. Consider two input creation operators $a_s^\dag$ and $a_i^\dag $ corresponding to signal and idler photons respectively. After a lossless, symmetric BS, the mode transformation for each component is given by:
$\left( {\begin{array}{*{20}{c}}
{a_s^\dag }\\
{a_i^\dag }
\end{array}} \right) = {U_{BS}}\left( {\begin{array}{*{20}{c}}
{a_1^\dag }\\
{a_2^\dag }
\end{array}} \right);{U_{BS}} = \left( {\begin{array}{*{20}{c}}
1&i\\
i&1
\end{array}} \right)$,
where ${a_1^\dag}$ and ${a_2^\dag}$ are the creation operators at the two output ports. Typically, photon sources for the quantum interference are generated via SPDC, the output two-photon state can be written as
\begin{eqnarray}
\left| \psi  \right\rangle &&= \int_0^\infty  {\int_0^\infty  {d{\omega _s}d{\omega _i}f({\omega _s},{\omega _i})} } a_s^\dag ({\omega _s})a_i^\dag ({\omega _i})\left| 0 \right\rangle,
 \nonumber\\
f({\omega _s},{\omega _i})&&= \alpha {E_P}({\omega _s} + {\omega _i})\sin c(\frac{{\Delta kL}}{2}){e^{i\phi ({\omega _s},{\omega _i})}}.
\end{eqnarray}
Here, $f({\omega _s},{\omega _i})$ is the joint spectral amplitude (JSA), describing the probability amplitude set by the pump envelope $E_P$ and the phase‑matching function. $\alpha$ is a normalization constant, and the convolution of $E_P$ with the sinc phase‑matching term enables generation of spectrally pure single photons. The phase mismatch for quasi‑phase matching in a crystal with poling period $\Lambda  $ is $\Delta k = {k_p}({\omega _s} + {\omega _i}) - {k_s}({\omega _s}) - {k_i}({\omega _i}) - \Lambda $, where each wavevector $k$ depends on its frequency and polarization. For convenience, we impose the normalized condition $\int_0^\infty  {\int_0^\infty  {d{\omega _1}d{\omega _2}} } |f({\omega _1},{\omega _2}){|^2} = 1$. In this work, we study two‑photon state generation under coherent and incoherent pumping.  Due to their differing phase correlations, the electric fields can be written as:
\begin{eqnarray}
{{\bf{E}}_C}(\omega ,t) &&= Af(x,y){\alpha _C}(\omega ){e^{ - i(\omega t - kz)}},
 \nonumber\\
{{\bf{E}}_I}(\omega ,t) &&= Af(x,y){\alpha _I}(\omega ){e^{ - i\varphi (\omega )}}{e^{ - i(\omega t - kz)}}.
\end{eqnarray}

In these expressions, $A$ is a constant and $f(x,y)$ is the normalized transverse spatial distribution describing the profile of the guided mode. The envelope functions ${\alpha _{C}}(\omega )$ and ${\alpha _{I}}(\omega )$ describe, respectively, the coherent and incoherent optical fields, with temporal incoherence in the latter introduced via a random phase factor $\phi (\omega )$.  Based on these definitions, the coincidence probability for HOM interference as a function of relative delay ${t_2} - {t_1} = \tau $ can be written as:
\begin{equation}
P(\tau ) = \int {\int {d{t_1}d{t_2}\left\langle \psi  \right|} } \hat E_1^{( - )}({t_1})\hat E_2^{( - )}({t_2})\hat E_2^{( + )}({t_2})\hat E_1^{( + )}({t_1})\left| \psi  \right\rangle, 
\end{equation}
where ${{\hat E}_1}$ and ${{\hat E}_2}$ are the field operators of detector 1 (D1) and detector 2 (D2) . After applying the BS transformation and integrating over time, this equation can be simplified as
\begin{equation}
P(\tau) =\frac{1}{4}\iint d\omega_1\,d\omega_2\; \bigl\lvert f(\omega_1,\omega_2) -\;f(\omega_2,\omega_1)\,e^{-i(\omega_1-\omega_2)\tau} \bigr\rvert^2.
\end{equation}
See more deduction details in Supplemental Materials and reference\cite{48}. For an ideal symmetric JSA satisfying ${f({\omega _1},{\omega _2}) = f({\omega _2},{\omega _1})}$, perfect photon bunching occurs at zero delay $\tau  = 0$. Based on the above results, HOM visibility is set exclusively by the overlap of the joint spectral intensities. Since every expression involves $f$ only through $|f{|^2}$, randomizing its phase does not alter the results. For second‑order intensity interference, phase difference between the photons cancels in the coincidence probability, so the visibility is governed solely by the indistinguishability of correlated photons.

In contrast, imperfect exchange symmetry ($f({\omega _1},{\omega _2}) \ne f({\omega _2},{\omega _1})$) prevents complete destructive interference, resulting in a nonzero residual coincidence background. In this scenario, JSA may be decomposed into its symmetric and antisymmetric components,${f_s}({\omega _1},{\omega _2}) = \frac{1}{2}[f({\omega _1},{\omega _2}) + f({\omega _2},{\omega _1})], {f_a}({\omega _1},{\omega _2}) = \frac{1}{2}[f({\omega _1},{\omega _2}) - f({\omega _2},{\omega _1})]$.Here ${f_s}$ accounts for the normalized perfectly symmetric part, and ${f_a}$ is the residual antisymmetric perturbation with total weight $\int_0^\infty  {\int_0^\infty  d } {\omega _1}d{\omega _2}|{f_a}({\omega _1},{\omega _2}){|^2} = \gamma $. Substituting into the coincidence‐probability formula, one finds in the idealized limit $P(0) = \gamma $. 

Two principal factors compromise photon indistinguishability. The first is polarization‑mode dispersion of the generated photons, which is determined by the crystal’s phase‑matching configuration. The second is spectral asymmetry of the pumping, which reduces the overlap of single‑photon wavepackets\cite{51,52}. Experimentally, beyond direct spectral or temporal filtering, SHG can serve as an effective nonlinear spectral filter. The cascaded SHG and SPDC approach has been widely adopted for pulse shaping and spectrum purification. For example, It has been applied in the generation of high-purity, heralded photons at 1550 nm for low-loss quantum key distribution\cite{50}.

Unlike SPDC, coherence is essential in the phase‑sensitive SHG process, previous studies have shown how incoherence increases conversion efficiency\cite{45}. Uncorrelated pump phases also affect the generated spectrum. Ignoring the phase‑matching envelope for clarity, we write the pumping amplitude as $A(\omega ) = {A_0}(\omega ) + \delta A(\omega )$, where ${A_0}$ the ideal symmetric profile about ${\omega _0}$ and $\delta A$ is a slowly varying perturbation. The generated fields then follow two distinct convolutions:
\begin{eqnarray}
{E_{{\rm{coh}}}}(\Omega ) &&= {\cal C}[A,A](\Omega ) = \int_{ - \infty }^{ + \infty } A (\omega )A(\Omega  - \omega ){\mkern 1mu} d\omega,\\
{E_{{\rm{inc}}}}(\Omega ) &&= \sqrt {{\cal C}[{\mkern 1mu} |A{|^2},{\mkern 1mu} |A{|^2}{\mkern 1mu} ](\Omega )}  = \sqrt {\int_{ - \infty }^{ + \infty } | A(\omega ){|^2}{\mkern 1mu} |A(\Omega  - \omega ){|^2}{\mkern 1mu} d\omega }.\nonumber
\end{eqnarray}

$E(\Omega )$ denotes the spectral amplitude of the generated field. The difference between the two expressions arises from the phase correlations among spectral components.  For incoherent light, the component phases are random, so $ < A({\omega _i})*A({\omega _j}) >  = 0,i \ne j$, and the process reduces to a self‑convolution of the intensity. In contrast, coherent light maintains consistent phase relationships, leading to a superposition of field amplitudes. This distinction between field convolution and intensity convolution determines each scheme’s sensitivity to a perturbation $\delta A$. Under a perturbative expansion $\delta A = \varepsilon {h_0}$, the ratio $R$ of the symmetric term to the first‑order perturbation at $\Omega  =2 {\omega _0}$ can be written as
\begin{equation}
{R_{{\rm{coh}}}} = 2\varepsilon {h_0} \cdot \frac{{\int {{A_0}} (\omega ){\mkern 1mu} d\omega }}{{\int {A_0^2} (\omega ){\mkern 1mu} d\omega }},{R_{{\rm{inc}}}} = 2\varepsilon {h_0} \cdot \frac{{\int {A_0^3} (\omega ){\mkern 1mu} d\omega }}{{\int {A_0^4} (\omega ){\mkern 1mu} d\omega }}.
\end{equation}
\begin{table}[htbp]
  \centering
  \caption{First‐order sensitivity coefficients for various pump lineshapes}
  \begin{tabular}{@{}lccc@{}}
    \hline
    Lineshape & $C_1={\int A_0\,d\omega}/{\int A_0^2\,d\omega}$ 
              & $C_2={\int A_0^3\,d\omega}/{\int A_0^4\,d\omega}$ 
              & $C_2/C_1$ \\
    \hline
    Gaussian               & $\sqrt{2}\approx1.414$ 
                           & $\frac{2}{\sqrt3}\approx1.155$ 
                           & $0.82$ \\
    Lorentzian             & $2$ 
                           & $1.2$ 
                           & $0.60$ \\
    Voigt (G  L)          & $1.039$ 
                           & $0.711$ 
                           & $0.68$ \\
    \hline
  \end{tabular}
\end{table}
See the Supplemental Materials for details. $\varepsilon$ is a small, dimensionless perturbation strength and the slowly varying shape function. The ratio between them depends primarily on the spectral line shape. Table 1 lists the $R$ values for three common spectral profiles. Under the same perturbation strength, incoherent illumination exhibits greater robustness to spectral asymmetry. Coherent SHG convolves field amplitudes, so tiny asymmetries in $A(\omega )$ are amplified by phase‑sensitive process. Incoherent SHG convolves intensities, averaging out phase correlations and yielding an output spectrum that preserves the original envelope under moderate perturbations.

\begin{figure}[hbtp]
\centering
\includegraphics[width=0.5\textwidth]{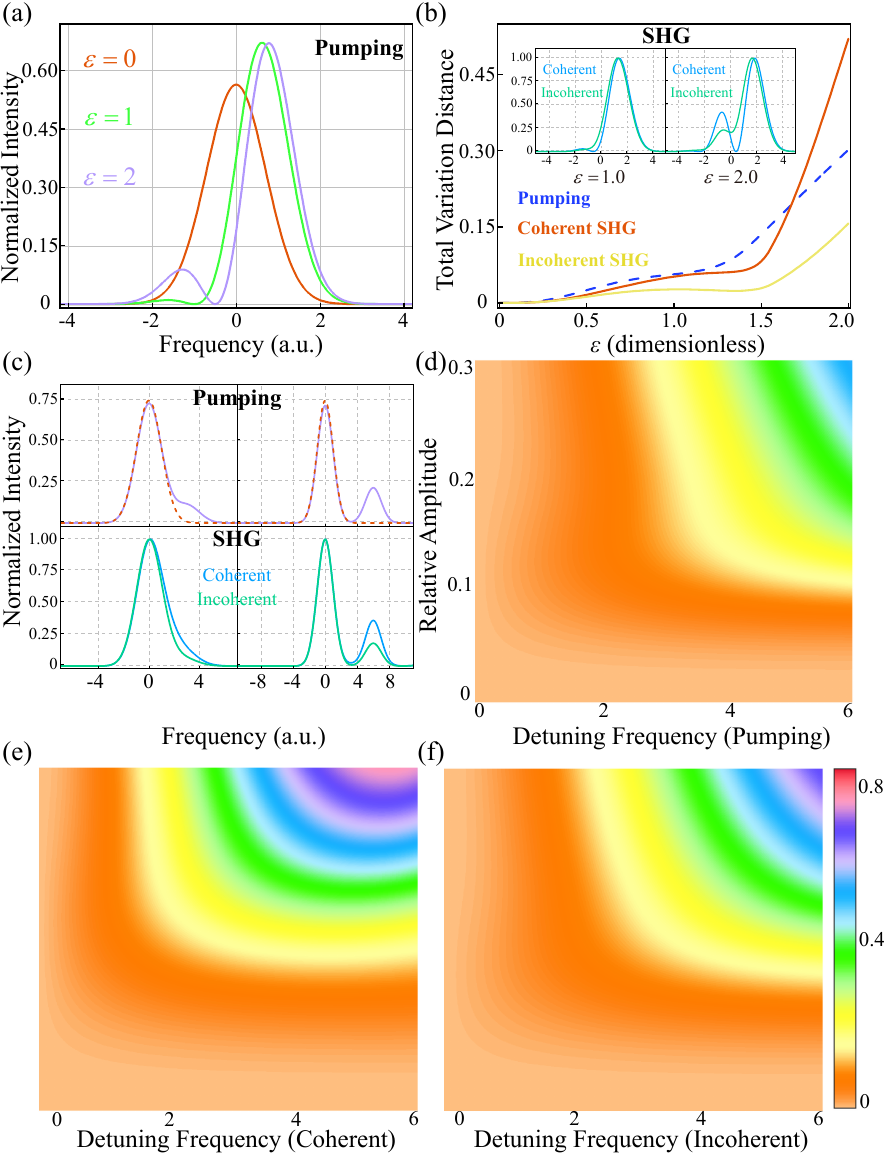}% Here is how to import EPS art
\caption{\label{fig:wide}
(a) Normalized pump intensity under linear spectral‑tilt perturbations of increasing strength. ;
(b)  TVD versus perturbation strength. Inset: SHG spectra generated by coherent and incoherent pumping at a given perturbation level.;
(c) Normalized intensity distributions of both pumping and SHG output under offset‑Gaussian superposition perturbations. Perturbations are described by amplitude ratio and spectral offset: left panel uses a ratio of 0.15 with an offset $b$ of $3\sigma $; right panel uses a ratio of 0.3 with b of $6\sigma $;
(d)-(f) TVD as functions of perturbation strength and spectral offset.}
\end{figure}

To highlight their differences, we carried out numerical simulations. Two complementary perturbation models are applied to capture the main forms of pumping asymmetry. Linear spectral tilt imposes a smooth slope across the entire bandwidth. The asymmetry often arises from residual gain tilt in broadband amplifiers or imperfect pulse compression. Offset Gaussian superposition adds a localized secondary peak or “satellite” pulse. The phenomenon is common when imperfect alignment or parasitic reflections introduce a displaced spectral component. Expressions for both models are given below, where $\varepsilon $ and $\omega$ denote dimensionless factors for numerical stimulation. Figures 1(a) and 1(c) illustrate the change in normalized intensity after applying the perturbations. Likewise, Figures 1(b) and 1(c) show the SHG spectra generated by coherent and incoherent pumping, respectively. In the incoherent case, the spectrum is noticeably smoother, since intensity convolution produces a more uniform energy distribution.
\begin{eqnarray}
{A_L}(\omega ) &&= {A_0}(\omega )(1 + \varepsilon {\mkern 1mu} \omega ),
 \nonumber\\
{A_G}(\omega ) &&= {A_0}[\exp ( - {\textstyle{{{{(\omega  - {\omega _0})}^2}} \over {2{\sigma ^2}}}}) + \varepsilon \;\exp ( - {\textstyle{{{{(\omega  - {\omega _0} - b\sigma )}^2}} \over {2{\sigma ^2}}}})].
\end{eqnarray}

Simultaneously, we employ the total variation distance (TVD) to quantify the spectral asymmetry of the generated field. It is defined as follows. $S(\omega ) = \frac{{|E(\omega ){|^2}}}{{\int | E(\omega ){|^2}{\mkern 1mu} d\omega }}$ is the normalized spectral intensity and $\mu  = \int {\omega S(\omega )d} \omega $ is the centroid. This metric directly measures the maximum deviation between a distribution and its mirror image\cite{53,54}. It is bounded between 0 and 1, enjoys a clear operational meaning, and captures all forms of asymmetry without bias toward particular moments. 
\begin{equation}
T = \frac{1}{2}\int_{ - \infty }^\infty  | {\mkern 1mu} S(\omega ) - S(2\mu  - \omega )|{\mkern 1mu} d\omega.
\end{equation}

Figure 1(b) and Figures 1(d)–1(f) show how spectral asymmetry changes with perturbation. We first calculated the pump’s asymmetry, shown by the dashed line in Figure 1(b) and in Figure 1(d). In the low‑perturbation region, the SHG output remains nearly symmetric. The generated field’s TVD grows more slowly than the pump’s. Under large spectral perturbations, coherent SHG’s field‑amplitude convolution amplifies the distorted pump profile, its asymmetry then grows rapidly and even surpasses the pump’s own, creating pronounced local peaks or troughs. In contrast, incoherent SHG’s intensity convolution simply adds the squared magnitudes, smoothing out irregularities and preserving the overall envelope shape even as the spectrum broadens.

\emph{Results.}-\begin{figure}[htbp]
\centering
\includegraphics[width=0.45\textwidth]{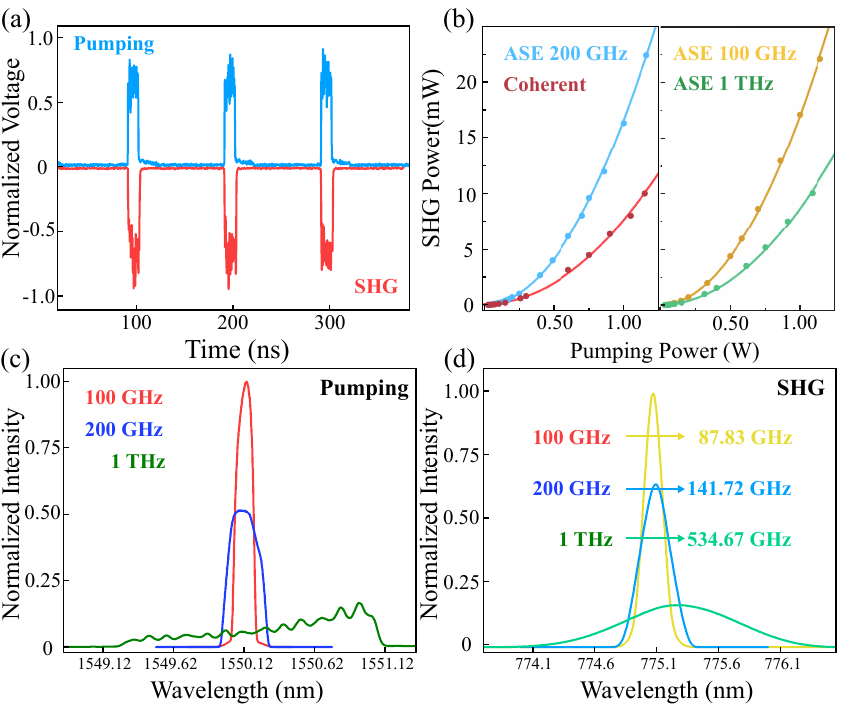}% Here is how to import EPS art
\caption{\label{fig:wide}
(a) Time‑domain profiles of the pumping and SHG fields;
(b) SHG power plotted against pump power, with quadratic fits applied. Both pumping coherence and spectral bandwidth critically limit conversion efficiency;
(c) Pumping spectral distribution. The envelope shape is governed primarily by the applied filter;
(d) SHG output spectrum. Its bandwidth is determined by the combined effects of pumping self‑convolution and the phase‑matching function;}
\end{figure}
In the experiment, we measure and characterize the key parameters and system properties described above. See the experimental setup in Supplemental Materials. Figure 2(a) shows the time-domain profiles of the 1550 nm pumping and the 775 nm SHG light. Although ASE noise is introduced during amplification, the pumping SNR can also reach 100. After SHG, the 775 nm output SNR exceeds 1000.

In the second-order nonlinear process, the conversion efficiency is governed by pumping coherence and phase-matching condition. Figure 2(b)  shows the SHG power versus pumping power. With the same level, the incoherent ASE source yields twice the SHG power of the coherent source. When the pumping bandwidth is 100 GHz or 200 GHz(both of which are narrower than the phase-matching bandwidth), the conversion efficiency remains unchanged. However, broadening the pump to 1 THz (using an 8 nm filter) introduces phase mismatch and significantly reduces efficiency.

Figures 2(d) and 2(e) illustrate how SHG reshapes and spectrally filters the envelope under incoherent pumping. As analyzed earlier, the intensity‑convolution process dramatically enhances spectral symmetry. From the conversion efficiency data in Figures 2(b) and 2(c), we see that for 100 GHz and 200 GHz pumping bandwidths, self‑convolution dominates. The TVD of pumping is 0.050 and 0.044, respectively, while that of the generated field is reduced to 0.004 and 0.008. At 500 GHz, the phase‑matching function not only limits the conversion bandwidth but also reshapes the generated spectral envelope. Despite a high pumping TVD of 0.414, the corresponding value for the generated field is suppressed to just 0.019. Consequently, the generated 775 nm spectra exhibit high symmetry in all cases.

Next, we characterize the quantum properties of the generated photon pairs. Figure 3(a) plots the coincidence counts and the coincidence-to-accidental ratio (CAR) versus pumping power. The coincidence count rate scales linearly with pumping power. At a peak power of 200 mW, the CAR remains above 13000. Error bars are based on Poissonian statistics. Because the type-II PPKTP crystal is only 2 mm long, the photon-pair spectrum spans more than 17 nm continuously.

\begin{figure}[h]
\centering
\includegraphics[width=0.5\textwidth]{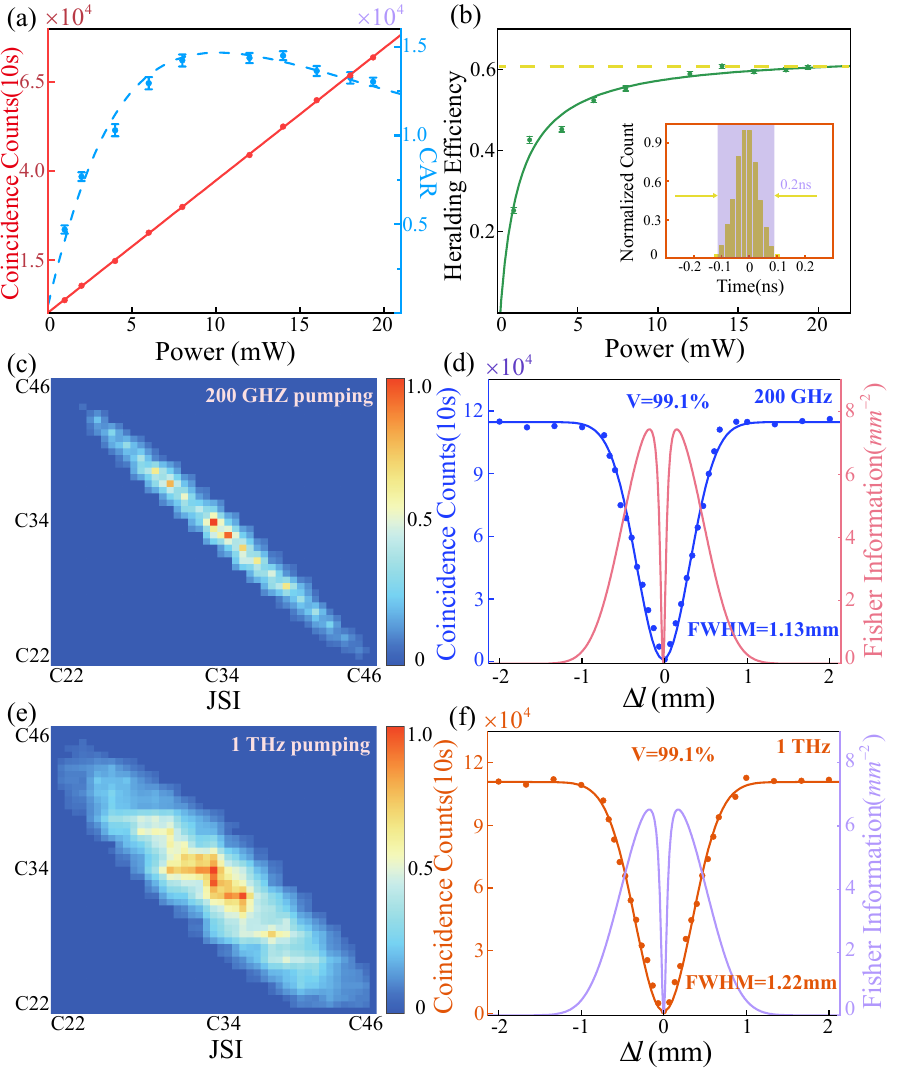}% Here is how to import EPS art
\caption{\label{fig:wide}
(a) Characterization of quantum correlation properties.;
(b) Characterization of the heralding efficiency. ;
(c) and (e) Joint spectral intensity for two pumping bandwidths.;
(d) and (f) HOM interference fringes and Fisher information as functions of relative delay. }
\end{figure}

The broadband photon distribution introduces lower temporal uncertainty, so the coincidence bin width is determined primarily by the system jitter. As shown in Figure 3(b), we calculate the signal-photon heralding efficiency from measured coincidence and single-side counts. The efficiency rises with pumping power and saturates near 60\%,  demonstrating high coupling efficiency and low loss of the system.

To analyze the spectral properties under incoherent pumping, we record the joint spectral intensity (JSI) and two‑photon HOM interference without spectral filtering. Figure 3(c) and 3(e) display the JSIs for 200 GHz and 1 THz pump bandwidths. In this experiment, the length of type-0 SHG PPKTP limits the maximum generated bandwidth. In the SPDC stage, the relaxed phase-matching condition produces a photon-pair spectrum wider than 17 nm, resulting in pronounced frequency anti-correlations. To generate high‑purity pairs directly, one can match the phase‑matching bandwidths of both ${\chi ^{(2)}}$  processes by optimizing crystal lengths.

Figures 3(d) and 3(f) present the HOM interference fringes for 200 GHz and 1 THz pumpings, respectively, both achieve 99.1\% visibility, confirming excellent spectral indistinguishability. For correlated photon pairs, the joint spectral distribution does not set the visibility. Instead, higher purity reduces spectral correlations and makes the two‑photon state more factorable. By the Fourier relation, weaker spectral correlations lead to broader temporal correlations. As a result, the HOM dip becomes wider, and the curve near zero delay becomes shallower. It directly cause the reducing sensitivity to small temporal shifts and thus lowering the Fisher information. The Fisher information $\mathcal{I}(\tau)$ quantifies how sharply the coincidence probability $P(\tau)$ changes with delay, it sets the lower bound on the variance of any unbiased estimator of $\tau$ via the Cramér–Rao bound. In HOM interferometry it is defined as ${\cal I}(\tau ) = \frac{{{{[{\partial _\tau }P(\tau )]}^2}}}{{P(\tau ){\mkern 1mu} [1 - P(\tau )]}}{\mkern 1mu}$\cite{55}. See more details in Supplemental Materials. Although high‑purity states benefit quantum state control, they inherently limit the precision of time‑delay estimation in HOM interferometry.

\emph{Discussion}--In conclusion, we analyze and demonstrate the creation of indistinguishable two photon states by incoherent pumping. Theoretically, HOM interference based on SPDC is immune to temporal incoherence, the randomly distributed pumping phase cancels out in second‑order intensity correlations, and the visibility of correlated source depends solely on the photon indistinguishability. Meanwhile, temporal incoherence converts the process into a intensity self‑convolution that averages out phase‑sensitive distortions in SHG. The generated spectra demonstrate robustness against perturbations and exhibit a highly symmetric profile.

In the experiment, the incoherence delivers a twofold enhancement in SHG conversion efficiency. The initially nonuniform pumping spectrum is converted into the spectrum with pronounced spectral symmetry. Using the incoherent second-harmonic light, we achieve the preparation of high-quality photon state through SPDC. The source achieves a CAR approaching 15000 at a coincidence count rate of 6.5 kHz. Even when the heralding efficiency maintains at 60\%, the CAR still exceeds 13000. The measured HOM interference fringes at different pump bandwidths exhibit a visibility of 99.1\% without any spectral filtering, confirming near‐ideal indistinguishability. Because the generated photons span a broad spectrum, changing the pumping bandwidth does not significantly affect visibility. However, variations in purity affect two-photon interference performance. Higher purity leads to a more factorable state, weakening spectral correlations. According to Fourier transform relations, this enhances temporal correlations, manifesting as broader HOM fringe bandwidth that ultimately limits interferometric estimation precision.

Our work establishes a practical route to generating high‑quality quantum states using readily available incoherent light sources. By demonstrating that phase randomness does not compromise HOM visibility and that intensity self‑convolution in SHG dramatically enhances spectral symmetry, we remove the need for expensive narrowband lasers and complex stabilization systems. Incoherent pumping such as ASE offer low cost, intrinsic robustness to environmental fluctuations, and relaxed stability requirements, making quantum states generation more accessible and scalable.

\begin{acknowledgments}
This work is supported by National Key Research and Development Program of China 2022YFB3903102,2022YFB3607700), National Natural Science Foundation of China (NSFC)(62435018), Innovation Program for Quantum Science and Technology (2021ZD0301100), USTC Research Funds of the Double First-Class Initiative (YD2030002023), and Research Cooperation Fund of SAST, CASC (SAST2022-075).
\end{acknowledgments}

\nocite{*}

\bibliography{Article}% Produces the bibliography via BibTeX.

\providecommand{\noopsort}[1]{}\providecommand{\singleletter}[1]{#1}%
\begin{thebibliography}{52}%
\makeatletter
\providecommand \@ifxundefined [1]{%
 \@ifx{#1\undefined}
}%
\providecommand \@ifnum [1]{%
 \ifnum #1\expandafter \@firstoftwo
 \else \expandafter \@secondoftwo
 \fi
}%
\providecommand \@ifx [1]{%
 \ifx #1\expandafter \@firstoftwo
 \else \expandafter \@secondoftwo
 \fi
}%
\providecommand \natexlab [1]{#1}%
\providecommand \enquote  [1]{``#1''}%
\providecommand \bibnamefont  [1]{#1}%
\providecommand \bibfnamefont [1]{#1}%
\providecommand \citenamefont [1]{#1}%
\providecommand \href@noop [0]{\@secondoftwo}%
\providecommand \href [0]{\begingroup \@sanitize@url \@href}%
\providecommand \@href[1]{\@@startlink{#1}\@@href}%
\providecommand \@@href[1]{\endgroup#1\@@endlink}%
\providecommand \@sanitize@url [0]{\catcode `\\12\catcode `\$12\catcode
  `\&12\catcode `\#12\catcode `\^12\catcode `\_12\catcode `\%12\relax}%
\providecommand \@@startlink[1]{}%
\providecommand \@@endlink[0]{}%
\providecommand \url  [0]{\begingroup\@sanitize@url \@url }%
\providecommand \@url [1]{\endgroup\@href {#1}{\urlprefix }}%
\providecommand \urlprefix  [0]{URL }%
\providecommand \Eprint [0]{\href }%
\providecommand \doibase [0]{https://doi.org/}%
\providecommand \selectlanguage [0]{\@gobble}%
\providecommand \bibinfo  [0]{\@secondoftwo}%
\providecommand \bibfield  [0]{\@secondoftwo}%
\providecommand \translation [1]{[#1]}%
\providecommand \BibitemOpen [0]{}%
\providecommand \bibitemStop [0]{}%
\providecommand \bibitemNoStop [0]{.\EOS\space}%
\providecommand \EOS [0]{\spacefactor3000\relax}%
\providecommand \BibitemShut  [1]{\csname bibitem#1\endcsname}%
\let\auto@bib@innerbib\@empty
%</preamble>
\bibitem [{\citenamefont {Scully}\ and\ \citenamefont {Zubairy}(1997)}]{1}%
  \BibitemOpen
  \bibfield  {author} {\bibinfo {author} {\bibfnamefont {M.~O.}\ \bibnamefont
  {Scully}}\ and\ \bibinfo {author} {\bibfnamefont {M.~S.}\ \bibnamefont
  {Zubairy}},\ }\href@noop {} {\emph {\bibinfo {title} {Quantum optics}}}\
  (\bibinfo  {publisher} {Cambridge University Press},\ \bibinfo {year}
  {1997})\BibitemShut {NoStop}%
\bibitem [{\citenamefont {Fabre}\ and\ \citenamefont {Treps}(2020)}]{2}%
  \BibitemOpen
  \bibfield  {author} {\bibinfo {author} {\bibfnamefont {C.}~\bibnamefont
  {Fabre}}\ and\ \bibinfo {author} {\bibfnamefont {N.}~\bibnamefont {Treps}},\
  }\bibfield  {title} {\bibinfo {title} {Modes and states in quantum optics},\
  }\href@noop {} {\bibfield  {journal} {\bibinfo  {journal} {Reviews of Modern
  Physics}\ }\textbf {\bibinfo {volume} {92}},\ \bibinfo {pages} {035005}
  (\bibinfo {year} {2020})}\BibitemShut {NoStop}%
\bibitem [{\citenamefont {Yamamoto}\ and\ \citenamefont {Haus}(1986)}]{3}%
  \BibitemOpen
  \bibfield  {author} {\bibinfo {author} {\bibfnamefont {Y.}~\bibnamefont
  {Yamamoto}}\ and\ \bibinfo {author} {\bibfnamefont {H.~A.}\ \bibnamefont
  {Haus}},\ }\bibfield  {title} {\bibinfo {title} {Preparation, measurement and
  information capacity of optical quantum states},\ }\href@noop {} {\bibfield
  {journal} {\bibinfo  {journal} {Reviews of Modern Physics}\ }\textbf
  {\bibinfo {volume} {58}},\ \bibinfo {pages} {1001} (\bibinfo {year}
  {1986})}\BibitemShut {NoStop}%
\bibitem [{\citenamefont {Dell’Anno}\ \emph {et~al.}(2006)\citenamefont
  {Dell’Anno}, \citenamefont {De~Siena},\ and\ \citenamefont
  {Illuminati}}]{4}%
  \BibitemOpen
  \bibfield  {author} {\bibinfo {author} {\bibfnamefont {F.}~\bibnamefont
  {Dell’Anno}}, \bibinfo {author} {\bibfnamefont {S.}~\bibnamefont
  {De~Siena}},\ and\ \bibinfo {author} {\bibfnamefont {F.}~\bibnamefont
  {Illuminati}},\ }\bibfield  {title} {\bibinfo {title} {Multiphoton quantum
  optics and quantum state engineering},\ }\href@noop {} {\bibfield  {journal}
  {\bibinfo  {journal} {Physics Reports}\ }\textbf {\bibinfo {volume} {428}},\
  \bibinfo {pages} {53} (\bibinfo {year} {2006})}\BibitemShut {NoStop}%
\bibitem [{\citenamefont {Caspani}\ \emph {et~al.}(2017)\citenamefont
  {Caspani}, \citenamefont {Xiong}, \citenamefont {Eggleton}, \citenamefont
  {Bajoni}, \citenamefont {Liscidini}, \citenamefont {Galli}, \citenamefont
  {Morandotti},\ and\ \citenamefont {Moss}}]{5}%
  \BibitemOpen
  \bibfield  {author} {\bibinfo {author} {\bibfnamefont {L.}~\bibnamefont
  {Caspani}}, \bibinfo {author} {\bibfnamefont {C.}~\bibnamefont {Xiong}},
  \bibinfo {author} {\bibfnamefont {B.~J.}\ \bibnamefont {Eggleton}}, \bibinfo
  {author} {\bibfnamefont {D.}~\bibnamefont {Bajoni}}, \bibinfo {author}
  {\bibfnamefont {M.}~\bibnamefont {Liscidini}}, \bibinfo {author}
  {\bibfnamefont {M.}~\bibnamefont {Galli}}, \bibinfo {author} {\bibfnamefont
  {R.}~\bibnamefont {Morandotti}},\ and\ \bibinfo {author} {\bibfnamefont
  {D.~J.}\ \bibnamefont {Moss}},\ }\bibfield  {title} {\bibinfo {title}
  {Integrated sources of photon quantum states based on nonlinear optics},\
  }\href@noop {} {\bibfield  {journal} {\bibinfo  {journal} {Light: Science \&
  Applications}\ }\textbf {\bibinfo {volume} {6}},\ \bibinfo {pages} {e17100}
  (\bibinfo {year} {2017})}\BibitemShut {NoStop}%
\bibitem [{\citenamefont {Paesani}\ \emph {et~al.}(2019)\citenamefont
  {Paesani}, \citenamefont {Ding}, \citenamefont {Santagati}, \citenamefont
  {Chakhmakhchyan}, \citenamefont {Vigliar}, \citenamefont {Rottwitt},
  \citenamefont {Oxenløwe}, \citenamefont {Wei}, \citenamefont {Thompson},\
  and\ \citenamefont {Laing}}]{6}%
  \BibitemOpen
  \bibfield  {author} {\bibinfo {author} {\bibfnamefont {S.}~\bibnamefont
  {Paesani}}, \bibinfo {author} {\bibfnamefont {Y.}~\bibnamefont {Ding}},
  \bibinfo {author} {\bibfnamefont {R.}~\bibnamefont {Santagati}}, \bibinfo
  {author} {\bibfnamefont {L.}~\bibnamefont {Chakhmakhchyan}}, \bibinfo
  {author} {\bibfnamefont {C.}~\bibnamefont {Vigliar}}, \bibinfo {author}
  {\bibfnamefont {K.}~\bibnamefont {Rottwitt}}, \bibinfo {author}
  {\bibfnamefont {L.}~\bibnamefont {Oxenløwe}}, \bibinfo {author}
  {\bibfnamefont {W.}~\bibnamefont {Wei}}, \bibinfo {author} {\bibfnamefont
  {M.}~\bibnamefont {Thompson}},\ and\ \bibinfo {author} {\bibfnamefont
  {A.}~\bibnamefont {Laing}},\ }\bibfield  {title} {\bibinfo {title}
  {Generation and sampling of quantum states of light in a silicon chip},\
  }\href@noop {} {\bibfield  {journal} {\bibinfo  {journal} {Nature Physics}\
  }\textbf {\bibinfo {volume} {15}},\ \bibinfo {pages} {925} (\bibinfo {year}
  {2019})}\BibitemShut {NoStop}%
\bibitem [{\citenamefont {Xu}\ \emph {et~al.}(2024)\citenamefont {Xu},
  \citenamefont {Su}, \citenamefont {Chai},\ and\ \citenamefont {Li}}]{7}%
  \BibitemOpen
  \bibfield  {author} {\bibinfo {author} {\bibfnamefont {Y.}~\bibnamefont
  {Xu}}, \bibinfo {author} {\bibfnamefont {X.}~\bibnamefont {Su}}, \bibinfo
  {author} {\bibfnamefont {Z.}~\bibnamefont {Chai}},\ and\ \bibinfo {author}
  {\bibfnamefont {J.}~\bibnamefont {Li}},\ }\bibfield  {title} {\bibinfo
  {title} {Metasurfaces toward optical manipulation technologies for quantum
  precision measurement},\ }\href@noop {} {\bibfield  {journal} {\bibinfo
  {journal} {Laser \& Photonics Reviews}\ }\textbf {\bibinfo {volume} {18}},\
  \bibinfo {pages} {2300355} (\bibinfo {year} {2024})}\BibitemShut {NoStop}%
\bibitem [{\citenamefont {Santori}\ \emph {et~al.}(2002)\citenamefont
  {Santori}, \citenamefont {Fattal}, \citenamefont {Vučković}, \citenamefont
  {Solomon},\ and\ \citenamefont {Yamamoto}}]{8}%
  \BibitemOpen
  \bibfield  {author} {\bibinfo {author} {\bibfnamefont {C.}~\bibnamefont
  {Santori}}, \bibinfo {author} {\bibfnamefont {D.}~\bibnamefont {Fattal}},
  \bibinfo {author} {\bibfnamefont {J.}~\bibnamefont {Vučković}}, \bibinfo
  {author} {\bibfnamefont {G.~S.}\ \bibnamefont {Solomon}},\ and\ \bibinfo
  {author} {\bibfnamefont {Y.}~\bibnamefont {Yamamoto}},\ }\bibfield  {title}
  {\bibinfo {title} {Indistinguishable photons from a single-photon device},\
  }\href@noop {} {\bibfield  {journal} {\bibinfo  {journal} {Nature}\ }\textbf
  {\bibinfo {volume} {419}},\ \bibinfo {pages} {594} (\bibinfo {year}
  {2002})}\BibitemShut {NoStop}%
\bibitem [{\citenamefont {Müller}\ \emph {et~al.}(2014)\citenamefont
  {Müller}, \citenamefont {Bounouar}, \citenamefont {Jöns}, \citenamefont
  {Glässl},\ and\ \citenamefont {Michler}}]{9}%
  \BibitemOpen
  \bibfield  {author} {\bibinfo {author} {\bibfnamefont {M.}~\bibnamefont
  {Müller}}, \bibinfo {author} {\bibfnamefont {S.}~\bibnamefont {Bounouar}},
  \bibinfo {author} {\bibfnamefont {K.~D.}\ \bibnamefont {Jöns}}, \bibinfo
  {author} {\bibfnamefont {M.}~\bibnamefont {Glässl}},\ and\ \bibinfo {author}
  {\bibfnamefont {P.}~\bibnamefont {Michler}},\ }\bibfield  {title} {\bibinfo
  {title} {On-demand generation of indistinguishable polarization-entangled
  photon pairs},\ }\href@noop {} {\bibfield  {journal} {\bibinfo  {journal}
  {Nature Photonics}\ }\textbf {\bibinfo {volume} {8}},\ \bibinfo {pages} {224}
  (\bibinfo {year} {2014})}\BibitemShut {NoStop}%
\bibitem [{\citenamefont {Hong}\ \emph {et~al.}(1987)\citenamefont {Hong},
  \citenamefont {Ou},\ and\ \citenamefont {Mandel}}]{10}%
  \BibitemOpen
  \bibfield  {author} {\bibinfo {author} {\bibfnamefont {C.~K.}\ \bibnamefont
  {Hong}}, \bibinfo {author} {\bibfnamefont {Z.~Y.}\ \bibnamefont {Ou}},\ and\
  \bibinfo {author} {\bibfnamefont {L.}~\bibnamefont {Mandel}},\ }\bibfield
  {title} {\bibinfo {title} {Measurement of subpicosecond time intervals
  between two photons by interference},\ }\href@noop {} {\bibfield  {journal}
  {\bibinfo  {journal} {Physical Review Letters}\ }\textbf {\bibinfo {volume}
  {59}},\ \bibinfo {pages} {2044} (\bibinfo {year} {1987})}\BibitemShut
  {NoStop}%
\bibitem [{\citenamefont {Khodadad~Kashi}\ \emph {et~al.}(2023)\citenamefont
  {Khodadad~Kashi}, \citenamefont {Caspani},\ and\ \citenamefont {Kues}}]{11}%
  \BibitemOpen
  \bibfield  {author} {\bibinfo {author} {\bibfnamefont {A.}~\bibnamefont
  {Khodadad~Kashi}}, \bibinfo {author} {\bibfnamefont {L.}~\bibnamefont
  {Caspani}},\ and\ \bibinfo {author} {\bibfnamefont {M.}~\bibnamefont
  {Kues}},\ }\bibfield  {title} {\bibinfo {title} {Spectral hong-ou-mandel
  effect between a heralded single-photon state and a thermal field:
  multiphoton contamination and the nonclassicality threshold},\ }\href@noop {}
  {\bibfield  {journal} {\bibinfo  {journal} {Physical Review Letters}\
  }\textbf {\bibinfo {volume} {131}},\ \bibinfo {pages} {233601} (\bibinfo
  {year} {2023})}\BibitemShut {NoStop}%
\bibitem [{\citenamefont {Ollivier}\ \emph {et~al.}(2021)\citenamefont
  {Ollivier}, \citenamefont {Thomas}, \citenamefont {Wein}, \citenamefont
  {Maillette~de Buy~Wenniger}, \citenamefont {Coste}, \citenamefont {Loredo},
  \citenamefont {Somaschi}, \citenamefont {Harouri}, \citenamefont {Lemaitre},
  \citenamefont {Sagnes}, \citenamefont {Lanco}, \citenamefont {Simon},
  \citenamefont {Anton}, \citenamefont {Krebs},\ and\ \citenamefont
  {Senellart}}]{12}%
  \BibitemOpen
  \bibfield  {author} {\bibinfo {author} {\bibfnamefont {H.}~\bibnamefont
  {Ollivier}}, \bibinfo {author} {\bibfnamefont {S.~E.}\ \bibnamefont
  {Thomas}}, \bibinfo {author} {\bibfnamefont {S.~C.}\ \bibnamefont {Wein}},
  \bibinfo {author} {\bibfnamefont {I.}~\bibnamefont {Maillette~de
  Buy~Wenniger}}, \bibinfo {author} {\bibfnamefont {N.}~\bibnamefont {Coste}},
  \bibinfo {author} {\bibfnamefont {J.~C.}\ \bibnamefont {Loredo}}, \bibinfo
  {author} {\bibfnamefont {N.}~\bibnamefont {Somaschi}}, \bibinfo {author}
  {\bibfnamefont {A.}~\bibnamefont {Harouri}}, \bibinfo {author} {\bibfnamefont
  {A.}~\bibnamefont {Lemaitre}}, \bibinfo {author} {\bibfnamefont
  {I.}~\bibnamefont {Sagnes}}, \bibinfo {author} {\bibfnamefont
  {L.}~\bibnamefont {Lanco}}, \bibinfo {author} {\bibfnamefont
  {C.}~\bibnamefont {Simon}}, \bibinfo {author} {\bibfnamefont
  {C.}~\bibnamefont {Anton}}, \bibinfo {author} {\bibfnamefont
  {O.}~\bibnamefont {Krebs}},\ and\ \bibinfo {author} {\bibfnamefont
  {P.}~\bibnamefont {Senellart}},\ }\bibfield  {title} {\bibinfo {title}
  {Hong-ou-mandel interference with imperfect single photon sources},\
  }\href@noop {} {\bibfield  {journal} {\bibinfo  {journal} {Physical Review
  Letters}\ }\textbf {\bibinfo {volume} {126}},\ \bibinfo {pages} {063602}
  (\bibinfo {year} {2021})}\BibitemShut {NoStop}%
\bibitem [{\citenamefont {Guo}\ \emph {et~al.}(2023)\citenamefont {Guo},
  \citenamefont {Yang}, \citenamefont {Zeng}, \citenamefont {Ding},
  \citenamefont {Shimizu},\ and\ \citenamefont {Jin}}]{13}%
  \BibitemOpen
  \bibfield  {author} {\bibinfo {author} {\bibfnamefont {Y.}~\bibnamefont
  {Guo}}, \bibinfo {author} {\bibfnamefont {Z.~X.}\ \bibnamefont {Yang}},
  \bibinfo {author} {\bibfnamefont {Z.~Q.}\ \bibnamefont {Zeng}}, \bibinfo
  {author} {\bibfnamefont {C.}~\bibnamefont {Ding}}, \bibinfo {author}
  {\bibfnamefont {R.}~\bibnamefont {Shimizu}},\ and\ \bibinfo {author}
  {\bibfnamefont {R.~B.}\ \bibnamefont {Jin}},\ }\bibfield  {title} {\bibinfo
  {title} {Comparison of multi-mode hong-ou-mandel interference and multi-slit
  interference},\ }\href@noop {} {\bibfield  {journal} {\bibinfo  {journal}
  {Optics Express}\ }\textbf {\bibinfo {volume} {31}},\ \bibinfo {pages}
  {32849} (\bibinfo {year} {2023})}\BibitemShut {NoStop}%
\bibitem [{\citenamefont {Kok}\ \emph {et~al.}(2007)\citenamefont {Kok},
  \citenamefont {Munro}, \citenamefont {Nemoto} \emph {et~al.}}]{14}%
  \BibitemOpen
  \bibfield  {author} {\bibinfo {author} {\bibfnamefont {P.}~\bibnamefont
  {Kok}}, \bibinfo {author} {\bibfnamefont {W.~J.}\ \bibnamefont {Munro}},
  \bibinfo {author} {\bibfnamefont {K.}~\bibnamefont {Nemoto}}, \emph
  {et~al.},\ }\bibfield  {title} {\bibinfo {title} {Linear optical quantum
  computing with photonic qubits},\ }\href@noop {} {\bibfield  {journal}
  {\bibinfo  {journal} {Reviews of Modern Physics}\ }\textbf {\bibinfo {volume}
  {79}},\ \bibinfo {pages} {135} (\bibinfo {year} {2007})}\BibitemShut
  {NoStop}%
\bibitem [{\citenamefont {Fabre}\ \emph {et~al.}(2022)\citenamefont {Fabre},
  \citenamefont {Amanti}, \citenamefont {Baboux}, \citenamefont {Keller},
  \citenamefont {Ducci},\ and\ \citenamefont {Milman}}]{15}%
  \BibitemOpen
  \bibfield  {author} {\bibinfo {author} {\bibfnamefont {N.}~\bibnamefont
  {Fabre}}, \bibinfo {author} {\bibfnamefont {M.}~\bibnamefont {Amanti}},
  \bibinfo {author} {\bibfnamefont {F.}~\bibnamefont {Baboux}}, \bibinfo
  {author} {\bibfnamefont {A.}~\bibnamefont {Keller}}, \bibinfo {author}
  {\bibfnamefont {S.}~\bibnamefont {Ducci}},\ and\ \bibinfo {author}
  {\bibfnamefont {P.}~\bibnamefont {Milman}},\ }\bibfield  {title} {\bibinfo
  {title} {The hong–ou–mandel experiment: from photon indistinguishability
  to continuous-variable quantum computing},\ }\href@noop {} {\bibfield
  {journal} {\bibinfo  {journal} {The European Physical Journal D}\ }\textbf
  {\bibinfo {volume} {76}},\ \bibinfo {pages} {196} (\bibinfo {year}
  {2022})}\BibitemShut {NoStop}%
\bibitem [{\citenamefont {Carolan}\ \emph {et~al.}(2015)\citenamefont
  {Carolan}, \citenamefont {Harrold}, \citenamefont {Sparrow}, \citenamefont
  {Martín-López}, \citenamefont {Russell}, \citenamefont {Silverstone},
  \citenamefont {Shadbolt}, \citenamefont {Matsuda}, \citenamefont {Oguma},
  \citenamefont {Itoh}, \citenamefont {Marshall}, \citenamefont {Thompson},
  \citenamefont {Matthews}, \citenamefont {Hashimoto}, \citenamefont
  {O’Brien},\ and\ \citenamefont {Laing}}]{16}%
  \BibitemOpen
  \bibfield  {author} {\bibinfo {author} {\bibfnamefont {J.}~\bibnamefont
  {Carolan}}, \bibinfo {author} {\bibfnamefont {C.}~\bibnamefont {Harrold}},
  \bibinfo {author} {\bibfnamefont {C.}~\bibnamefont {Sparrow}}, \bibinfo
  {author} {\bibfnamefont {E.}~\bibnamefont {Martín-López}}, \bibinfo
  {author} {\bibfnamefont {N.~J.}\ \bibnamefont {Russell}}, \bibinfo {author}
  {\bibfnamefont {J.~W.}\ \bibnamefont {Silverstone}}, \bibinfo {author}
  {\bibfnamefont {P.~J.}\ \bibnamefont {Shadbolt}}, \bibinfo {author}
  {\bibfnamefont {N.}~\bibnamefont {Matsuda}}, \bibinfo {author} {\bibfnamefont
  {M.}~\bibnamefont {Oguma}}, \bibinfo {author} {\bibfnamefont
  {M.}~\bibnamefont {Itoh}}, \bibinfo {author} {\bibfnamefont {G.~D.}\
  \bibnamefont {Marshall}}, \bibinfo {author} {\bibfnamefont {M.~G.}\
  \bibnamefont {Thompson}}, \bibinfo {author} {\bibfnamefont {J.~C.~F.}\
  \bibnamefont {Matthews}}, \bibinfo {author} {\bibfnamefont {T.}~\bibnamefont
  {Hashimoto}}, \bibinfo {author} {\bibfnamefont {J.~L.}\ \bibnamefont
  {O’Brien}},\ and\ \bibinfo {author} {\bibfnamefont {A.}~\bibnamefont
  {Laing}},\ }\bibfield  {title} {\bibinfo {title} {Universal linear optics},\
  }\href@noop {} {\bibfield  {journal} {\bibinfo  {journal} {Science}\ }\textbf
  {\bibinfo {volume} {349}},\ \bibinfo {pages} {711} (\bibinfo {year}
  {2015})}\BibitemShut {NoStop}%
\bibitem [{\citenamefont {Lo}\ \emph {et~al.}(2012)\citenamefont {Lo},
  \citenamefont {Curty},\ and\ \citenamefont {Qi}}]{17}%
  \BibitemOpen
  \bibfield  {author} {\bibinfo {author} {\bibfnamefont {H.~K.}\ \bibnamefont
  {Lo}}, \bibinfo {author} {\bibfnamefont {M.}~\bibnamefont {Curty}},\ and\
  \bibinfo {author} {\bibfnamefont {B.}~\bibnamefont {Qi}},\ }\bibfield
  {title} {\bibinfo {title} {Measurement-device-independent quantum key
  distribution},\ }\href@noop {} {\bibfield  {journal} {\bibinfo  {journal}
  {Physical Review Letters}\ }\textbf {\bibinfo {volume} {108}},\ \bibinfo
  {pages} {130503} (\bibinfo {year} {2012})}\BibitemShut {NoStop}%
\bibitem [{\citenamefont {Zhan}\ \emph
  {et~al.}(2025{\natexlab{a}})\citenamefont {Zhan}, \citenamefont {Zhong},
  \citenamefont {Ma}, \citenamefont {Wang}, \citenamefont {Yin}, \citenamefont
  {Chen}, \citenamefont {He}, \citenamefont {Guo},\ and\ \citenamefont
  {Han}}]{18}%
  \BibitemOpen
  \bibfield  {author} {\bibinfo {author} {\bibfnamefont {X.~H.}\ \bibnamefont
  {Zhan}}, \bibinfo {author} {\bibfnamefont {Z.~Q.}\ \bibnamefont {Zhong}},
  \bibinfo {author} {\bibfnamefont {J.~Y.}\ \bibnamefont {Ma}}, \bibinfo
  {author} {\bibfnamefont {S.}~\bibnamefont {Wang}}, \bibinfo {author}
  {\bibfnamefont {Z.~Q.}\ \bibnamefont {Yin}}, \bibinfo {author} {\bibfnamefont
  {W.}~\bibnamefont {Chen}}, \bibinfo {author} {\bibfnamefont {D.~Y.}\
  \bibnamefont {He}}, \bibinfo {author} {\bibfnamefont {G.~C.}\ \bibnamefont
  {Guo}},\ and\ \bibinfo {author} {\bibfnamefont {Z.~F.}\ \bibnamefont {Han}},\
  }\bibfield  {title} {\bibinfo {title} {Experimental demonstration of long
  distance quantum communication with independent heralded single photon
  sources},\ }\href@noop {} {\bibfield  {journal} {\bibinfo  {journal} {npj
  Quantum Information}\ }\textbf {\bibinfo {volume} {11}},\ \bibinfo {pages}
  {1} (\bibinfo {year} {2025}{\natexlab{a}})}\BibitemShut {NoStop}%
\bibitem [{\citenamefont {Wang}\ \emph {et~al.}(2024)\citenamefont {Wang},
  \citenamefont {Lu}, \citenamefont {Wang}, \citenamefont {Yin}, \citenamefont
  {Wang}, \citenamefont {Geng}, \citenamefont {Chen}, \citenamefont {He},
  \citenamefont {Guo},\ and\ \citenamefont {Han}}]{19}%
  \BibitemOpen
  \bibfield  {author} {\bibinfo {author} {\bibfnamefont {X.}~\bibnamefont
  {Wang}}, \bibinfo {author} {\bibfnamefont {F.~Y.}\ \bibnamefont {Lu}},
  \bibinfo {author} {\bibfnamefont {Z.~H.}\ \bibnamefont {Wang}}, \bibinfo
  {author} {\bibfnamefont {Z.~Q.}\ \bibnamefont {Yin}}, \bibinfo {author}
  {\bibfnamefont {S.}~\bibnamefont {Wang}}, \bibinfo {author} {\bibfnamefont
  {J.~Q.}\ \bibnamefont {Geng}}, \bibinfo {author} {\bibfnamefont
  {W.}~\bibnamefont {Chen}}, \bibinfo {author} {\bibfnamefont {D.~Y.}\
  \bibnamefont {He}}, \bibinfo {author} {\bibfnamefont {G.~C.}\ \bibnamefont
  {Guo}},\ and\ \bibinfo {author} {\bibfnamefont {Z.~F.}\ \bibnamefont {Han}},\
  }\bibfield  {title} {\bibinfo {title} {Fully passive
  measurement-device-independent quantum key distribution},\ }\href@noop {}
  {\bibfield  {journal} {\bibinfo  {journal} {Physical Review Applied}\
  }\textbf {\bibinfo {volume} {21}},\ \bibinfo {pages} {064067} (\bibinfo
  {year} {2024})}\BibitemShut {NoStop}%
\bibitem [{\citenamefont {Lyons}\ \emph {et~al.}(2018)\citenamefont {Lyons},
  \citenamefont {Knee}, \citenamefont {Bolduc}, \citenamefont {Roger},
  \citenamefont {Leach}, \citenamefont {Gauger},\ and\ \citenamefont
  {Faccio}}]{20}%
  \BibitemOpen
  \bibfield  {author} {\bibinfo {author} {\bibfnamefont {A.}~\bibnamefont
  {Lyons}}, \bibinfo {author} {\bibfnamefont {G.~C.}\ \bibnamefont {Knee}},
  \bibinfo {author} {\bibfnamefont {E.}~\bibnamefont {Bolduc}}, \bibinfo
  {author} {\bibfnamefont {T.}~\bibnamefont {Roger}}, \bibinfo {author}
  {\bibfnamefont {J.}~\bibnamefont {Leach}}, \bibinfo {author} {\bibfnamefont
  {E.~M.}\ \bibnamefont {Gauger}},\ and\ \bibinfo {author} {\bibfnamefont
  {D.}~\bibnamefont {Faccio}},\ }\bibfield  {title} {\bibinfo {title}
  {Attosecond-resolution hong–ou–mandel interferometry},\ }\href@noop {}
  {\bibfield  {journal} {\bibinfo  {journal} {Science Advances}\ }\textbf
  {\bibinfo {volume} {4}},\ \bibinfo {pages} {eaap9416} (\bibinfo {year}
  {2018})}\BibitemShut {NoStop}%
\bibitem [{\citenamefont {Singh}\ \emph {et~al.}(2024)\citenamefont {Singh},
  \citenamefont {Kumar},\ and\ \citenamefont {Samanta}}]{21}%
  \BibitemOpen
  \bibfield  {author} {\bibinfo {author} {\bibfnamefont {S.}~\bibnamefont
  {Singh}}, \bibinfo {author} {\bibfnamefont {V.}~\bibnamefont {Kumar}},\ and\
  \bibinfo {author} {\bibfnamefont {G.~K.}\ \bibnamefont {Samanta}},\
  }\bibfield  {title} {\bibinfo {title} {Fast measurement of group index
  variation with optimum precision using hong–ou–mandel interferometry},\
  }\href@noop {} {\bibfield  {journal} {\bibinfo  {journal} {APL Quantum}\
  }\textbf {\bibinfo {volume} {1}} (\bibinfo {year} {2024})}\BibitemShut
  {NoStop}%
\bibitem [{\citenamefont {Boyd}\ \emph {et~al.}(2008)\citenamefont {Boyd},
  \citenamefont {Gaeta},\ and\ \citenamefont {Giese}}]{22}%
  \BibitemOpen
  \bibfield  {author} {\bibinfo {author} {\bibfnamefont {R.~W.}\ \bibnamefont
  {Boyd}}, \bibinfo {author} {\bibfnamefont {A.~L.}\ \bibnamefont {Gaeta}},\
  and\ \bibinfo {author} {\bibfnamefont {E.}~\bibnamefont {Giese}},\ }\bibfield
   {title} {\bibinfo {title} {Nonlinear optics},\ }in\ \href@noop {} {\emph
  {\bibinfo {booktitle} {Springer Handbook of Atomic, Molecular, and Optical
  Physics}}}\ (\bibinfo  {publisher} {Springer},\ \bibinfo {year} {2008})\ pp.\
  \bibinfo {pages} {1097--1110}\BibitemShut {NoStop}%
\bibitem [{\citenamefont {Couteau}(2018)}]{23}%
  \BibitemOpen
  \bibfield  {author} {\bibinfo {author} {\bibfnamefont {C.}~\bibnamefont
  {Couteau}},\ }\bibfield  {title} {\bibinfo {title} {Spontaneous parametric
  down-conversion},\ }\href@noop {} {\bibfield  {journal} {\bibinfo  {journal}
  {Contemporary Physics}\ }\textbf {\bibinfo {volume} {59}},\ \bibinfo {pages}
  {291} (\bibinfo {year} {2018})}\BibitemShut {NoStop}%
\bibitem [{\citenamefont {Zhang}\ \emph {et~al.}(2021)\citenamefont {Zhang},
  \citenamefont {Huang}, \citenamefont {Liu}, \citenamefont {Li},\ and\
  \citenamefont {Guo}}]{24}%
  \BibitemOpen
  \bibfield  {author} {\bibinfo {author} {\bibfnamefont {C.}~\bibnamefont
  {Zhang}}, \bibinfo {author} {\bibfnamefont {Y.~F.}\ \bibnamefont {Huang}},
  \bibinfo {author} {\bibfnamefont {B.~H.}\ \bibnamefont {Liu}}, \bibinfo
  {author} {\bibfnamefont {C.~F.}\ \bibnamefont {Li}},\ and\ \bibinfo {author}
  {\bibfnamefont {G.~C.}\ \bibnamefont {Guo}},\ }\bibfield  {title} {\bibinfo
  {title} {Spontaneous parametric down‐conversion sources for multiphoton
  experiments},\ }\href@noop {} {\bibfield  {journal} {\bibinfo  {journal}
  {Advanced Quantum Technologies}\ }\textbf {\bibinfo {volume} {4}},\ \bibinfo
  {pages} {2000132} (\bibinfo {year} {2021})}\BibitemShut {NoStop}%
\bibitem [{\citenamefont {Cundiff}\ and\ \citenamefont {Ye}(2005)}]{56}%
  \BibitemOpen
  \bibfield  {author} {\bibinfo {author} {\bibfnamefont {S.~T.}\ \bibnamefont
  {Cundiff}}\ and\ \bibinfo {author} {\bibfnamefont {J.}~\bibnamefont {Ye}},\
  }\bibfield  {title} {\bibinfo {title} {Phase stabilization of mode-locked
  lasers},\ }\href {https://doi.org/10.1080/09500340412331313704} {\bibfield
  {journal} {\bibinfo  {journal} {Journal of Modern Optics}\ }\textbf {\bibinfo
  {volume} {52}},\ \bibinfo {pages} {201} (\bibinfo {year} {2005})}\BibitemShut
  {NoStop}%
\bibitem [{\citenamefont {Bayer}\ \emph {et~al.}(2021)\citenamefont {Bayer},
  \citenamefont {Li}, \citenamefont {Guentchev}, \citenamefont {Torun},
  \citenamefont {Velazco},\ and\ \citenamefont {Boyraz}}]{29}%
  \BibitemOpen
  \bibfield  {author} {\bibinfo {author} {\bibfnamefont {M.~M.}\ \bibnamefont
  {Bayer}}, \bibinfo {author} {\bibfnamefont {X.}~\bibnamefont {Li}}, \bibinfo
  {author} {\bibfnamefont {G.~N.}\ \bibnamefont {Guentchev}}, \bibinfo {author}
  {\bibfnamefont {R.}~\bibnamefont {Torun}}, \bibinfo {author} {\bibfnamefont
  {J.~E.}\ \bibnamefont {Velazco}},\ and\ \bibinfo {author} {\bibfnamefont
  {O.}~\bibnamefont {Boyraz}},\ }\bibfield  {title} {\bibinfo {title}
  {Single-shot ranging and velocimetry with a cw lidar far beyond the coherence
  length of the cw laser},\ }\href@noop {} {\bibfield  {journal} {\bibinfo
  {journal} {Optics Express}\ }\textbf {\bibinfo {volume} {29}},\ \bibinfo
  {pages} {42343} (\bibinfo {year} {2021})}\BibitemShut {NoStop}%
\bibitem [{\citenamefont {Gisin}\ and\ \citenamefont {Thew}(2007)}]{30}%
  \BibitemOpen
  \bibfield  {author} {\bibinfo {author} {\bibfnamefont {N.}~\bibnamefont
  {Gisin}}\ and\ \bibinfo {author} {\bibfnamefont {R.}~\bibnamefont {Thew}},\
  }\bibfield  {title} {\bibinfo {title} {Quantum communication},\ }\href@noop
  {} {\bibfield  {journal} {\bibinfo  {journal} {Nature Photonics}\ }\textbf
  {\bibinfo {volume} {1}},\ \bibinfo {pages} {165} (\bibinfo {year}
  {2007})}\BibitemShut {NoStop}%
\bibitem [{\citenamefont {Giovannetti}\ \emph {et~al.}(2004)\citenamefont
  {Giovannetti}, \citenamefont {Lloyd},\ and\ \citenamefont {Maccone}}]{31}%
  \BibitemOpen
  \bibfield  {author} {\bibinfo {author} {\bibfnamefont {V.}~\bibnamefont
  {Giovannetti}}, \bibinfo {author} {\bibfnamefont {S.}~\bibnamefont {Lloyd}},\
  and\ \bibinfo {author} {\bibfnamefont {L.}~\bibnamefont {Maccone}},\
  }\bibfield  {title} {\bibinfo {title} {Quantum-enhanced measurements: beating
  the standard quantum limit},\ }\href@noop {} {\bibfield  {journal} {\bibinfo
  {journal} {Science}\ }\textbf {\bibinfo {volume} {306}},\ \bibinfo {pages}
  {1330} (\bibinfo {year} {2004})}\BibitemShut {NoStop}%
\bibitem [{\citenamefont {Arahira}\ and\ \citenamefont
  {Murai}(2014{\natexlab{a}})}]{32}%
  \BibitemOpen
  \bibfield  {author} {\bibinfo {author} {\bibfnamefont {S.}~\bibnamefont
  {Arahira}}\ and\ \bibinfo {author} {\bibfnamefont {H.}~\bibnamefont
  {Murai}},\ }\bibfield  {title} {\bibinfo {title} {Wavelength conversion of
  incoherent light by sum-frequency generation},\ }\href@noop {} {\bibfield
  {journal} {\bibinfo  {journal} {Optics Express}\ }\textbf {\bibinfo {volume}
  {22}},\ \bibinfo {pages} {12944} (\bibinfo {year}
  {2014}{\natexlab{a}})}\BibitemShut {NoStop}%
\bibitem [{\citenamefont {Giese}\ \emph {et~al.}(2018)\citenamefont {Giese},
  \citenamefont {Fickler}, \citenamefont {Zhang}, \citenamefont {Chen},\ and\
  \citenamefont {Boyd}}]{33}%
  \BibitemOpen
  \bibfield  {author} {\bibinfo {author} {\bibfnamefont {E.}~\bibnamefont
  {Giese}}, \bibinfo {author} {\bibfnamefont {R.}~\bibnamefont {Fickler}},
  \bibinfo {author} {\bibfnamefont {W.}~\bibnamefont {Zhang}}, \bibinfo
  {author} {\bibfnamefont {L.}~\bibnamefont {Chen}},\ and\ \bibinfo {author}
  {\bibfnamefont {R.~W.}\ \bibnamefont {Boyd}},\ }\bibfield  {title} {\bibinfo
  {title} {Influence of pump coherence on the quantum properties of spontaneous
  parametric down-conversion},\ }\href@noop {} {\bibfield  {journal} {\bibinfo
  {journal} {Physica Scripta}\ }\textbf {\bibinfo {volume} {93}},\ \bibinfo
  {pages} {084001} (\bibinfo {year} {2018})}\BibitemShut {NoStop}%
\bibitem [{\citenamefont {Zhao}\ \emph {et~al.}(2020)\citenamefont {Zhao},
  \citenamefont {Ji}, \citenamefont {Liu}, \citenamefont {Gao}, \citenamefont
  {Rao}, \citenamefont {Cui}, \citenamefont {Feng}, \citenamefont {Li},
  \citenamefont {Shi}, \citenamefont {Shan}, \citenamefont {Ma},\ and\
  \citenamefont {Sui}}]{34}%
  \BibitemOpen
  \bibfield  {author} {\bibinfo {author} {\bibfnamefont {X.}~\bibnamefont
  {Zhao}}, \bibinfo {author} {\bibfnamefont {L.}~\bibnamefont {Ji}}, \bibinfo
  {author} {\bibfnamefont {D.}~\bibnamefont {Liu}}, \bibinfo {author}
  {\bibfnamefont {Y.}~\bibnamefont {Gao}}, \bibinfo {author} {\bibfnamefont
  {D.}~\bibnamefont {Rao}}, \bibinfo {author} {\bibfnamefont {Y.}~\bibnamefont
  {Cui}}, \bibinfo {author} {\bibfnamefont {W.}~\bibnamefont {Feng}}, \bibinfo
  {author} {\bibfnamefont {F.}~\bibnamefont {Li}}, \bibinfo {author}
  {\bibfnamefont {H.}~\bibnamefont {Shi}}, \bibinfo {author} {\bibfnamefont
  {C.}~\bibnamefont {Shan}}, \bibinfo {author} {\bibfnamefont {W.}~\bibnamefont
  {Ma}},\ and\ \bibinfo {author} {\bibfnamefont {Z.}~\bibnamefont {Sui}},\
  }\bibfield  {title} {\bibinfo {title} {Second-harmonic generation of
  temporally low-coherence light},\ }\href@noop {} {\bibfield  {journal}
  {\bibinfo  {journal} {APL Photonics}\ }\textbf {\bibinfo {volume} {5}}
  (\bibinfo {year} {2020})}\BibitemShut {NoStop}%
\bibitem [{\citenamefont {Hutter}\ \emph {et~al.}(2020)\citenamefont {Hutter},
  \citenamefont {Lima},\ and\ \citenamefont {Walborn}}]{35}%
  \BibitemOpen
  \bibfield  {author} {\bibinfo {author} {\bibfnamefont {L.}~\bibnamefont
  {Hutter}}, \bibinfo {author} {\bibfnamefont {G.}~\bibnamefont {Lima}},\ and\
  \bibinfo {author} {\bibfnamefont {S.~P.}\ \bibnamefont {Walborn}},\
  }\bibfield  {title} {\bibinfo {title} {Boosting entanglement generation in
  down-conversion with incoherent illumination},\ }\href@noop {} {\bibfield
  {journal} {\bibinfo  {journal} {Physical Review Letters}\ }\textbf {\bibinfo
  {volume} {125}},\ \bibinfo {pages} {193602} (\bibinfo {year}
  {2020})}\BibitemShut {NoStop}%
\bibitem [{\citenamefont {Tan}\ \emph {et~al.}(2023{\natexlab{a}})\citenamefont
  {Tan}, \citenamefont {Qi}, \citenamefont {Chen}, \citenamefont {Danner},
  \citenamefont {Kanchanawong},\ and\ \citenamefont {Tsang}}]{36}%
  \BibitemOpen
  \bibfield  {author} {\bibinfo {author} {\bibfnamefont {X.~J.}\ \bibnamefont
  {Tan}}, \bibinfo {author} {\bibfnamefont {L.}~\bibnamefont {Qi}}, \bibinfo
  {author} {\bibfnamefont {L.}~\bibnamefont {Chen}}, \bibinfo {author}
  {\bibfnamefont {A.~J.}\ \bibnamefont {Danner}}, \bibinfo {author}
  {\bibfnamefont {P.}~\bibnamefont {Kanchanawong}},\ and\ \bibinfo {author}
  {\bibfnamefont {M.}~\bibnamefont {Tsang}},\ }\bibfield  {title} {\bibinfo
  {title} {Quantum-inspired superresolution for incoherent imaging},\
  }\href@noop {} {\bibfield  {journal} {\bibinfo  {journal} {Optica}\ }\textbf
  {\bibinfo {volume} {10}},\ \bibinfo {pages} {1189} (\bibinfo {year}
  {2023}{\natexlab{a}})}\BibitemShut {NoStop}%
\bibitem [{\citenamefont {Lupo}\ \emph {et~al.}(2020)\citenamefont {Lupo},
  \citenamefont {Huang},\ and\ \citenamefont {Kok}}]{37}%
  \BibitemOpen
  \bibfield  {author} {\bibinfo {author} {\bibfnamefont {C.}~\bibnamefont
  {Lupo}}, \bibinfo {author} {\bibfnamefont {Z.}~\bibnamefont {Huang}},\ and\
  \bibinfo {author} {\bibfnamefont {P.}~\bibnamefont {Kok}},\ }\bibfield
  {title} {\bibinfo {title} {Quantum limits to incoherent imaging are achieved
  by linear interferometry},\ }\href@noop {} {\bibfield  {journal} {\bibinfo
  {journal} {Physical Review Letters}\ }\textbf {\bibinfo {volume} {124}},\
  \bibinfo {pages} {080503} (\bibinfo {year} {2020})}\BibitemShut {NoStop}%
\bibitem [{\citenamefont {Dong}\ \emph {et~al.}(2024)\citenamefont {Dong},
  \citenamefont {Brückerhoff-Plückelmann}, \citenamefont {Meyer},
  \citenamefont {Dijkstra}, \citenamefont {Bente}, \citenamefont {Wendland},
  \citenamefont {Varri}, \citenamefont {Aggarwal}, \citenamefont {Farmakidis},
  \citenamefont {Wang}, \citenamefont {Yang}, \citenamefont {Lee},
  \citenamefont {He}, \citenamefont {Gooskens}, \citenamefont {Kwong},
  \citenamefont {Bienstman}, \citenamefont {Pernice},\ and\ \citenamefont
  {Bhaskaran}}]{38}%
  \BibitemOpen
  \bibfield  {author} {\bibinfo {author} {\bibfnamefont {B.}~\bibnamefont
  {Dong}}, \bibinfo {author} {\bibfnamefont {F.}~\bibnamefont
  {Brückerhoff-Plückelmann}}, \bibinfo {author} {\bibfnamefont
  {L.}~\bibnamefont {Meyer}}, \bibinfo {author} {\bibfnamefont
  {J.}~\bibnamefont {Dijkstra}}, \bibinfo {author} {\bibfnamefont
  {I.}~\bibnamefont {Bente}}, \bibinfo {author} {\bibfnamefont
  {D.}~\bibnamefont {Wendland}}, \bibinfo {author} {\bibfnamefont
  {A.}~\bibnamefont {Varri}}, \bibinfo {author} {\bibfnamefont
  {S.}~\bibnamefont {Aggarwal}}, \bibinfo {author} {\bibfnamefont
  {N.}~\bibnamefont {Farmakidis}}, \bibinfo {author} {\bibfnamefont
  {M.}~\bibnamefont {Wang}}, \bibinfo {author} {\bibfnamefont {G.}~\bibnamefont
  {Yang}}, \bibinfo {author} {\bibfnamefont {J.~S.}\ \bibnamefont {Lee}},
  \bibinfo {author} {\bibfnamefont {Y.}~\bibnamefont {He}}, \bibinfo {author}
  {\bibfnamefont {E.}~\bibnamefont {Gooskens}}, \bibinfo {author}
  {\bibfnamefont {D.-L.}\ \bibnamefont {Kwong}}, \bibinfo {author}
  {\bibfnamefont {P.}~\bibnamefont {Bienstman}}, \bibinfo {author}
  {\bibfnamefont {W.~H.~P.}\ \bibnamefont {Pernice}},\ and\ \bibinfo {author}
  {\bibfnamefont {H.}~\bibnamefont {Bhaskaran}},\ }\bibfield  {title} {\bibinfo
  {title} {Partial coherence enhances parallelized photonic computing},\
  }\href@noop {} {\bibfield  {journal} {\bibinfo  {journal} {Nature}\ }\textbf
  {\bibinfo {volume} {632}},\ \bibinfo {pages} {55} (\bibinfo {year}
  {2024})}\BibitemShut {NoStop}%
\bibitem [{\citenamefont {Song}\ \emph
  {et~al.}(2024{\natexlab{a}})\citenamefont {Song}, \citenamefont
  {Murty~Kottapalli}, \citenamefont {Goyal}, \citenamefont {Schölkopf},\ and\
  \citenamefont {Fischer}}]{39}%
  \BibitemOpen
  \bibfield  {author} {\bibinfo {author} {\bibfnamefont {A.}~\bibnamefont
  {Song}}, \bibinfo {author} {\bibfnamefont {S.~N.}\ \bibnamefont
  {Murty~Kottapalli}}, \bibinfo {author} {\bibfnamefont {R.}~\bibnamefont
  {Goyal}}, \bibinfo {author} {\bibfnamefont {B.}~\bibnamefont {Schölkopf}},\
  and\ \bibinfo {author} {\bibfnamefont {P.}~\bibnamefont {Fischer}},\
  }\bibfield  {title} {\bibinfo {title} {Low-power scalable multilayer
  optoelectronic neural networks enabled with incoherent light},\ }\href@noop
  {} {\bibfield  {journal} {\bibinfo  {journal} {Nature Communications}\
  }\textbf {\bibinfo {volume} {15}},\ \bibinfo {pages} {10692} (\bibinfo {year}
  {2024}{\natexlab{a}})}\BibitemShut {NoStop}%
\bibitem [{\citenamefont {Lee}\ \emph {et~al.}(2023)\citenamefont {Lee},
  \citenamefont {Kim}, \citenamefont {Im}, \citenamefont {Kim}, \citenamefont
  {Tamma},\ and\ \citenamefont {Kim}}]{40}%
  \BibitemOpen
  \bibfield  {author} {\bibinfo {author} {\bibfnamefont {C.~H.}\ \bibnamefont
  {Lee}}, \bibinfo {author} {\bibfnamefont {Y.}~\bibnamefont {Kim}}, \bibinfo
  {author} {\bibfnamefont {D.~G.}\ \bibnamefont {Im}}, \bibinfo {author}
  {\bibfnamefont {U.~S.}\ \bibnamefont {Kim}}, \bibinfo {author} {\bibfnamefont
  {V.}~\bibnamefont {Tamma}},\ and\ \bibinfo {author} {\bibfnamefont {Y.~H.}\
  \bibnamefont {Kim}},\ }\bibfield  {title} {\bibinfo {title} {Coherent
  two-photon lidar with incoherent light},\ }\href@noop {} {\bibfield
  {journal} {\bibinfo  {journal} {Physical Review Letters}\ }\textbf {\bibinfo
  {volume} {131}},\ \bibinfo {pages} {223602} (\bibinfo {year}
  {2023})}\BibitemShut {NoStop}%
\bibitem [{\citenamefont {Tan}\ \emph {et~al.}(2023{\natexlab{b}})\citenamefont
  {Tan}, \citenamefont {Yeo}, \citenamefont {Leow}, \citenamefont {Shen},\ and\
  \citenamefont {Kurtsiefer}}]{41}%
  \BibitemOpen
  \bibfield  {author} {\bibinfo {author} {\bibfnamefont {P.~K.}\ \bibnamefont
  {Tan}}, \bibinfo {author} {\bibfnamefont {X.~J.}\ \bibnamefont {Yeo}},
  \bibinfo {author} {\bibfnamefont {A.~Z.~W.}\ \bibnamefont {Leow}}, \bibinfo
  {author} {\bibfnamefont {L.}~\bibnamefont {Shen}},\ and\ \bibinfo {author}
  {\bibfnamefont {C.}~\bibnamefont {Kurtsiefer}},\ }\bibfield  {title}
  {\bibinfo {title} {Practical range sensing with thermal light},\ }\href@noop
  {} {\bibfield  {journal} {\bibinfo  {journal} {Physical Review Applied}\
  }\textbf {\bibinfo {volume} {20}},\ \bibinfo {pages} {014060} (\bibinfo
  {year} {2023}{\natexlab{b}})}\BibitemShut {NoStop}%
\bibitem [{\citenamefont {Li}\ \emph {et~al.}(2023)\citenamefont {Li},
  \citenamefont {Braverman}, \citenamefont {Kulkarni},\ and\ \citenamefont
  {Boyd}}]{42}%
  \BibitemOpen
  \bibfield  {author} {\bibinfo {author} {\bibfnamefont {C.}~\bibnamefont
  {Li}}, \bibinfo {author} {\bibfnamefont {B.}~\bibnamefont {Braverman}},
  \bibinfo {author} {\bibfnamefont {G.}~\bibnamefont {Kulkarni}},\ and\
  \bibinfo {author} {\bibfnamefont {R.~W.}\ \bibnamefont {Boyd}},\ }\bibfield
  {title} {\bibinfo {title} {Experimental generation of polarization
  entanglement from spontaneous parametric down-conversion pumped by
  spatiotemporally highly incoherent light},\ }\href@noop {} {\bibfield
  {journal} {\bibinfo  {journal} {Physical Review A}\ }\textbf {\bibinfo
  {volume} {107}},\ \bibinfo {pages} {L041701} (\bibinfo {year}
  {2023})}\BibitemShut {NoStop}%
\bibitem [{\citenamefont {Zhang}\ \emph {et~al.}(2023)\citenamefont {Zhang},
  \citenamefont {Xu},\ and\ \citenamefont {Chen}}]{43}%
  \BibitemOpen
  \bibfield  {author} {\bibinfo {author} {\bibfnamefont {W.}~\bibnamefont
  {Zhang}}, \bibinfo {author} {\bibfnamefont {D.}~\bibnamefont {Xu}},\ and\
  \bibinfo {author} {\bibfnamefont {L.}~\bibnamefont {Chen}},\ }\bibfield
  {title} {\bibinfo {title} {Polarization entanglement from parametric
  down-conversion with an led pump},\ }\href@noop {} {\bibfield  {journal}
  {\bibinfo  {journal} {Physical Review Applied}\ }\textbf {\bibinfo {volume}
  {19}},\ \bibinfo {pages} {054079} (\bibinfo {year} {2023})}\BibitemShut
  {NoStop}%
\bibitem [{\citenamefont {Song}\ \emph
  {et~al.}(2024{\natexlab{b}})\citenamefont {Song}, \citenamefont {Zhao},
  \citenamefont {Chen}, \citenamefont {Li}, \citenamefont {Li}, \citenamefont
  {Gao}, \citenamefont {Chen}, \citenamefont {Han}, \citenamefont {Xie},
  \citenamefont {Guo}, \citenamefont {Zhou},\ and\ \citenamefont {Shi}}]{44}%
  \BibitemOpen
  \bibfield  {author} {\bibinfo {author} {\bibfnamefont {Y.~W.}\ \bibnamefont
  {Song}}, \bibinfo {author} {\bibfnamefont {H.}~\bibnamefont {Zhao}}, \bibinfo
  {author} {\bibfnamefont {L.}~\bibnamefont {Chen}}, \bibinfo {author}
  {\bibfnamefont {Y.~H.}\ \bibnamefont {Li}}, \bibinfo {author} {\bibfnamefont
  {E.~Z.}\ \bibnamefont {Li}}, \bibinfo {author} {\bibfnamefont {M.~Y.}\
  \bibnamefont {Gao}}, \bibinfo {author} {\bibfnamefont {R.~H.}\ \bibnamefont
  {Chen}}, \bibinfo {author} {\bibfnamefont {Z.~Q.~Z.}\ \bibnamefont {Han}},
  \bibinfo {author} {\bibfnamefont {M.~Y.}\ \bibnamefont {Xie}}, \bibinfo
  {author} {\bibfnamefont {G.~C.}\ \bibnamefont {Guo}}, \bibinfo {author}
  {\bibfnamefont {Z.~Y.}\ \bibnamefont {Zhou}},\ and\ \bibinfo {author}
  {\bibfnamefont {B.~S.}\ \bibnamefont {Shi}},\ }\href@noop {} {\bibinfo
  {title} {On-chip quantum states generation by incoherent light}},\ \bibinfo
  {howpublished} {arXiv preprint arXiv:2412.03802} (\bibinfo {year}
  {2024}{\natexlab{b}})\BibitemShut {NoStop}%
\bibitem [{\citenamefont {Plöschner}\ \emph {et~al.}(2022)\citenamefont
  {Plöschner}, \citenamefont {Morote}, \citenamefont {Dahl}, \citenamefont
  {Mounaix}, \citenamefont {Light}, \citenamefont {Rakić},\ and\ \citenamefont
  {Carpenter}}]{57}%
  \BibitemOpen
  \bibfield  {author} {\bibinfo {author} {\bibfnamefont {M.}~\bibnamefont
  {Plöschner}}, \bibinfo {author} {\bibfnamefont {M.~M.}\ \bibnamefont
  {Morote}}, \bibinfo {author} {\bibfnamefont {D.~S.}\ \bibnamefont {Dahl}},
  \bibinfo {author} {\bibfnamefont {M.}~\bibnamefont {Mounaix}}, \bibinfo
  {author} {\bibfnamefont {G.}~\bibnamefont {Light}}, \bibinfo {author}
  {\bibfnamefont {A.~D.}\ \bibnamefont {Rakić}},\ and\ \bibinfo {author}
  {\bibfnamefont {J.}~\bibnamefont {Carpenter}},\ }\bibfield  {title} {\bibinfo
  {title} {Spatial tomography of light resolved in time, spectrum, and
  polarisation},\ }\href {https://doi.org/10.1038/s41467-022-31968-2}
  {\bibfield  {journal} {\bibinfo  {journal} {Nature Communications}\ }\textbf
  {\bibinfo {volume} {13}},\ \bibinfo {pages} {4294} (\bibinfo {year}
  {2022})}\BibitemShut {NoStop}%
\bibitem [{\citenamefont {Arahira}\ and\ \citenamefont
  {Murai}(2014{\natexlab{b}})}]{45}%
  \BibitemOpen
  \bibfield  {author} {\bibinfo {author} {\bibfnamefont {S.}~\bibnamefont
  {Arahira}}\ and\ \bibinfo {author} {\bibfnamefont {H.}~\bibnamefont
  {Murai}},\ }\bibfield  {title} {\bibinfo {title} {Wavelength conversion of
  incoherent light by sum-frequency generation},\ }\href@noop {} {\bibfield
  {journal} {\bibinfo  {journal} {Optics Express}\ }\textbf {\bibinfo {volume}
  {22}},\ \bibinfo {pages} {12944} (\bibinfo {year}
  {2014}{\natexlab{b}})}\BibitemShut {NoStop}%
\bibitem [{\citenamefont {Doronin}\ \emph {et~al.}(2019)\citenamefont
  {Doronin}, \citenamefont {Andrianov}, \citenamefont {Zyablovsky},
  \citenamefont {Pukhov}, \citenamefont {Lozovik}, \citenamefont {Vinogradov},\
  and\ \citenamefont {Lisyansky}}]{46}%
  \BibitemOpen
  \bibfield  {author} {\bibinfo {author} {\bibfnamefont {I.~V.}\ \bibnamefont
  {Doronin}}, \bibinfo {author} {\bibfnamefont {E.~S.}\ \bibnamefont
  {Andrianov}}, \bibinfo {author} {\bibfnamefont {A.~A.}\ \bibnamefont
  {Zyablovsky}}, \bibinfo {author} {\bibfnamefont {A.~A.}\ \bibnamefont
  {Pukhov}}, \bibinfo {author} {\bibfnamefont {Y.~E.}\ \bibnamefont {Lozovik}},
  \bibinfo {author} {\bibfnamefont {A.~P.}\ \bibnamefont {Vinogradov}},\ and\
  \bibinfo {author} {\bibfnamefont {A.~A.}\ \bibnamefont {Lisyansky}},\
  }\bibfield  {title} {\bibinfo {title} {Second-order coherence properties of
  amplified spontaneous emission},\ }\href@noop {} {\bibfield  {journal}
  {\bibinfo  {journal} {Optics Express}\ }\textbf {\bibinfo {volume} {27}},\
  \bibinfo {pages} {10991} (\bibinfo {year} {2019})}\BibitemShut {NoStop}%
\bibitem [{\citenamefont {Valero}\ \emph {et~al.}(2021)\citenamefont {Valero},
  \citenamefont {Marion}, \citenamefont {Lhermite}, \citenamefont {Delagnes},
  \citenamefont {Renard}, \citenamefont {Royon},\ and\ \citenamefont
  {Cormier}}]{47}%
  \BibitemOpen
  \bibfield  {author} {\bibinfo {author} {\bibfnamefont {N.}~\bibnamefont
  {Valero}}, \bibinfo {author} {\bibfnamefont {D.}~\bibnamefont {Marion}},
  \bibinfo {author} {\bibfnamefont {J.}~\bibnamefont {Lhermite}}, \bibinfo
  {author} {\bibfnamefont {J.~C.}\ \bibnamefont {Delagnes}}, \bibinfo {author}
  {\bibfnamefont {W.}~\bibnamefont {Renard}}, \bibinfo {author} {\bibfnamefont
  {R.}~\bibnamefont {Royon}},\ and\ \bibinfo {author} {\bibfnamefont
  {E.}~\bibnamefont {Cormier}},\ }\bibfield  {title} {\bibinfo {title}
  {High-power amplified spontaneous emission pulses with tunable coherence for
  efficient non-linear processes},\ }\href@noop {} {\bibfield  {journal}
  {\bibinfo  {journal} {Scientific Reports}\ }\textbf {\bibinfo {volume}
  {11}},\ \bibinfo {pages} {4844} (\bibinfo {year} {2021})}\BibitemShut
  {NoStop}%
\bibitem [{\citenamefont {Jin}\ \emph {et~al.}(2024)\citenamefont {Jin},
  \citenamefont {Zeng}, \citenamefont {You},\ and\ \citenamefont {Yuan}}]{48}%
  \BibitemOpen
  \bibfield  {author} {\bibinfo {author} {\bibfnamefont {R.~B.}\ \bibnamefont
  {Jin}}, \bibinfo {author} {\bibfnamefont {Z.~Q.}\ \bibnamefont {Zeng}},
  \bibinfo {author} {\bibfnamefont {C.}~\bibnamefont {You}},\ and\ \bibinfo
  {author} {\bibfnamefont {C.}~\bibnamefont {Yuan}},\ }\bibfield  {title}
  {\bibinfo {title} {Quantum interferometers: principles and applications},\
  }\href@noop {} {\bibfield  {journal} {\bibinfo  {journal} {Progress in
  Quantum Electronics}\ ,\ \bibinfo {pages} {100519}} (\bibinfo {year}
  {2024})}\BibitemShut {NoStop}%
\bibitem [{\citenamefont {Arzani}\ \emph {et~al.}(2018)\citenamefont {Arzani},
  \citenamefont {Fabre},\ and\ \citenamefont {Treps}}]{51}%
  \BibitemOpen
  \bibfield  {author} {\bibinfo {author} {\bibfnamefont {F.}~\bibnamefont
  {Arzani}}, \bibinfo {author} {\bibfnamefont {C.}~\bibnamefont {Fabre}},\ and\
  \bibinfo {author} {\bibfnamefont {N.}~\bibnamefont {Treps}},\ }\bibfield
  {title} {\bibinfo {title} {Versatile engineering of multimode squeezed states
  by optimizing the pump spectral profile in spontaneous parametric
  down-conversion},\ }\href@noop {} {\bibfield  {journal} {\bibinfo  {journal}
  {Physical Review A}\ }\textbf {\bibinfo {volume} {97}},\ \bibinfo {pages}
  {033808} (\bibinfo {year} {2018})}\BibitemShut {NoStop}%
\bibitem [{\citenamefont {Kaneda}\ \emph {et~al.}(2020)\citenamefont {Kaneda},
  \citenamefont {Oikawa}, \citenamefont {Yabuno}, \citenamefont {China},
  \citenamefont {Miki}, \citenamefont {Terai}, \citenamefont {Mitsumori},\ and\
  \citenamefont {Edamatsu}}]{52}%
  \BibitemOpen
  \bibfield  {author} {\bibinfo {author} {\bibfnamefont {F.}~\bibnamefont
  {Kaneda}}, \bibinfo {author} {\bibfnamefont {J.}~\bibnamefont {Oikawa}},
  \bibinfo {author} {\bibfnamefont {M.}~\bibnamefont {Yabuno}}, \bibinfo
  {author} {\bibfnamefont {F.}~\bibnamefont {China}}, \bibinfo {author}
  {\bibfnamefont {S.}~\bibnamefont {Miki}}, \bibinfo {author} {\bibfnamefont
  {H.}~\bibnamefont {Terai}}, \bibinfo {author} {\bibfnamefont
  {Y.}~\bibnamefont {Mitsumori}},\ and\ \bibinfo {author} {\bibfnamefont
  {K.}~\bibnamefont {Edamatsu}},\ }\bibfield  {title} {\bibinfo {title}
  {Spectral characterization of photon-pair sources via classical sum-frequency
  generation},\ }\href@noop {} {\bibfield  {journal} {\bibinfo  {journal}
  {Optics Express}\ }\textbf {\bibinfo {volume} {28}},\ \bibinfo {pages}
  {38993} (\bibinfo {year} {2020})}\BibitemShut {NoStop}%
\bibitem [{\citenamefont {Zhan}\ \emph
  {et~al.}(2025{\natexlab{b}})\citenamefont {Zhan}, \citenamefont {Zhong},
  \citenamefont {Ma}, \citenamefont {Wang}, \citenamefont {Yin}, \citenamefont
  {Chen}, \citenamefont {He}, \citenamefont {Guo},\ and\ \citenamefont
  {Han}}]{50}%
  \BibitemOpen
  \bibfield  {author} {\bibinfo {author} {\bibfnamefont {X.~H.}\ \bibnamefont
  {Zhan}}, \bibinfo {author} {\bibfnamefont {Z.~Q.}\ \bibnamefont {Zhong}},
  \bibinfo {author} {\bibfnamefont {J.~Y.}\ \bibnamefont {Ma}}, \bibinfo
  {author} {\bibfnamefont {S.}~\bibnamefont {Wang}}, \bibinfo {author}
  {\bibfnamefont {Z.~Q.}\ \bibnamefont {Yin}}, \bibinfo {author} {\bibfnamefont
  {W.}~\bibnamefont {Chen}}, \bibinfo {author} {\bibfnamefont {D.~Y.}\
  \bibnamefont {He}}, \bibinfo {author} {\bibfnamefont {G.~C.}\ \bibnamefont
  {Guo}},\ and\ \bibinfo {author} {\bibfnamefont {Z.~F.}\ \bibnamefont {Han}},\
  }\bibfield  {title} {\bibinfo {title} {Experimental demonstration of long
  distance quantum communication with independent heralded single photon
  sources},\ }\href@noop {} {\bibfield  {journal} {\bibinfo  {journal} {npj
  Quantum Information}\ }\textbf {\bibinfo {volume} {11}},\ \bibinfo {pages}
  {1} (\bibinfo {year} {2025}{\natexlab{b}})}\BibitemShut {NoStop}%
\bibitem [{\citenamefont {Fuchs}\ and\ \citenamefont {Van
  De~Graaf}(2002)}]{53}%
  \BibitemOpen
  \bibfield  {author} {\bibinfo {author} {\bibfnamefont {C.~A.}\ \bibnamefont
  {Fuchs}}\ and\ \bibinfo {author} {\bibfnamefont {J.}~\bibnamefont {Van
  De~Graaf}},\ }\bibfield  {title} {\bibinfo {title} {Cryptographic
  distinguishability measures for quantum-mechanical states},\ }\href@noop {}
  {\bibfield  {journal} {\bibinfo  {journal} {IEEE Transactions on Information
  Theory}\ }\textbf {\bibinfo {volume} {45}},\ \bibinfo {pages} {1216}
  (\bibinfo {year} {2002})}\BibitemShut {NoStop}%
\bibitem [{\citenamefont {Gross}\ \emph {et~al.}(2010)\citenamefont {Gross},
  \citenamefont {Liu}, \citenamefont {Flammia}, \citenamefont {Becker},\ and\
  \citenamefont {Eisert}}]{54}%
  \BibitemOpen
  \bibfield  {author} {\bibinfo {author} {\bibfnamefont {D.}~\bibnamefont
  {Gross}}, \bibinfo {author} {\bibfnamefont {Y.~K.}\ \bibnamefont {Liu}},
  \bibinfo {author} {\bibfnamefont {S.~T.}\ \bibnamefont {Flammia}}, \bibinfo
  {author} {\bibfnamefont {S.}~\bibnamefont {Becker}},\ and\ \bibinfo {author}
  {\bibfnamefont {J.}~\bibnamefont {Eisert}},\ }\bibfield  {title} {\bibinfo
  {title} {Quantum state tomography via compressed sensing},\ }\href@noop {}
  {\bibfield  {journal} {\bibinfo  {journal} {Physical Review Letters}\
  }\textbf {\bibinfo {volume} {105}},\ \bibinfo {pages} {150401} (\bibinfo
  {year} {2010})}\BibitemShut {NoStop}%
\bibitem [{\citenamefont {Braunstein}\ and\ \citenamefont {Caves}(1994)}]{55}%
  \BibitemOpen
  \bibfield  {author} {\bibinfo {author} {\bibfnamefont {S.~L.}\ \bibnamefont
  {Braunstein}}\ and\ \bibinfo {author} {\bibfnamefont {C.~M.}\ \bibnamefont
  {Caves}},\ }\bibfield  {title} {\bibinfo {title} {Statistical distance and
  the geometry of quantum states},\ }\href@noop {} {\bibfield  {journal}
  {\bibinfo  {journal} {Physical Review Letters}\ }\textbf {\bibinfo {volume}
  {72}},\ \bibinfo {pages} {3439} (\bibinfo {year} {1994})}\BibitemShut
  {NoStop}%
\end{thebibliography}%

\end{document}